\documentclass[preprint,12pt]{elsarticle}
\usepackage{epsfig}
\usepackage{array}
\usepackage{amssymb}
\usepackage{placeins}
\usepackage{amsmath}
\usepackage{amsthm}
\usepackage{url}
\usepackage{booktabs}
\usepackage{multirow, multicol}
\usepackage{adjustbox}
\usepackage{float}
\usepackage[table]{xcolor}
\usepackage[dvipsnames]{xcolor}
\usepackage{booktabs}
\usepackage{subcaption}
\usepackage{multirow}
\usepackage{comment}
\usepackage{algorithm}
\usepackage{algpseudocode}
\usepackage{esvect}
\usepackage{tikz}
\usepackage{graphicx}
\newcolumntype{M}[1]{>{\centering\arraybackslash}m{#1}} 
\newcolumntype{P}[1]{>{\raggedright\arraybackslash}p{#1}}

\definecolor{DesignBlue}{HTML}{8DA0CB}
\definecolor{DesignOrange}{HTML}{FC8D62}
\definecolor{SoftGreen}{HTML}{c5e1a5}

\theoremstyle{definition}

\definecolor{lightblue}{HTML}{A2DFF7}%
\definecolor{lightpink}{rgb}{1.0, 0.75, 0.8}
\definecolor{lightgreen}{rgb}{0.564, 0.933, 0.564}

\journal{TDB}

\begin{document}

\begin{frontmatter}
\title{An Optimal Battery-Free Approach for Emission Reduction by Storing Solar Surplus in Building Thermal Mass} 

\author[label1]{Michela Boffi\corref{cor1}}
\ead{michela.boffi@polimi.it}
\cortext[cor1]{Corresponding author}
\author[label1]{Jessica Leoni}
\ead{jessica.leoni@polimi.it}
\author[label1]{Fabrizio Leonforte}
\ead{fabrizio.leonforte@polimi.it}
\author[label1]{Mara Tanelli}
\ead{mara.tanelli@polimi.it}
\author[label1]{Paolo Oliaro}
\ead{paolo.oliaro@polimi.it}

\affiliation[label1]{organization={Politecnico di Milano},
            addressline={Piazza Leonardo da Vinci},
            city={Milano},
            postcode={20133},
            state={Milano},
            country={Italy}}
 
%% Abstract
\begin{abstract}
Decarbonization in buildings calls for advanced control strategies that coordinate on-site renewables, grid electricity, and thermal demand. Literature approaches typically rely on demand side management strategies or on active energy storage, like batteries. However, the first solution often neglects carbon-aware objectives, and could lead to grid overload issues, while batteries entail environmental, end-of-life, and cost concerns.
\newline To overcome these limitations, we propose an optimal, carbon-aware optimization strategy that exploits the building’s thermal mass as a passive storage, avoiding dedicated batteries. Specifically, when a surplus of renewable energy is available, our strategy computes the optimal share of surplus to store by temporarily adjusting the indoor temperature setpoint within comfort bounds. Thus, by explicitly accounting for forecasts of building energy consumption, solar production, and time-varying grid carbon intensity, our strategy enables emissions-aware load shifting while maintaining comfort. We evaluate the approach by simulating three TRNSYS models of the same system with different thermal mass. In all cases, the results show consistent reductions in grid electricity consumption with respect to a baseline that does not leverage surplus renewable generation. These findings highlight the potential of thermal-mass-based control for building decarbonization.
\end{abstract}


\end{frontmatter}

%% Sections
\section{Introduction}
\label{sec:intro}
The construction and building sector contributes significantly to global energy-related $\text{CO}_2$ emissions. According to data reported in the Global Status Report for Buildings and Construction 2024/2025, in 2023 buildings were responsible for approximately 32\% of global energy demand and 34\% of $\text{CO}_2$ emissions, mainly associated with space heating and cooling, domestic hot water production, lighting, and other end uses, which reached a record of 9.8 Gt$\text{CO}_2$~\cite{UNEPGlobalABC2024BeyondFoundations}.
Despite the progress made in 2024, represented by greater adoption of renewable energy and increased electrification, the sector is still not aligned with the climate neutrality targets set for 2050. In fact, as highlighted by the global buildings climate tracker, between 2015 and 2023, $\text{CO}_2$ emissions associated with buildings increased by 5.4\%, compared to the required reduction of 28.1\%~\cite{UNEPGlobalABC2024BeyondFoundations}.
As a consequence, the observed progress remains insufficient to meet the targets of the Paris Agreement~\cite{UNEPGlobalABC2024BeyondFoundations}.

\begin{figure}[h!]
	\centering
	\includegraphics[width=\linewidth]{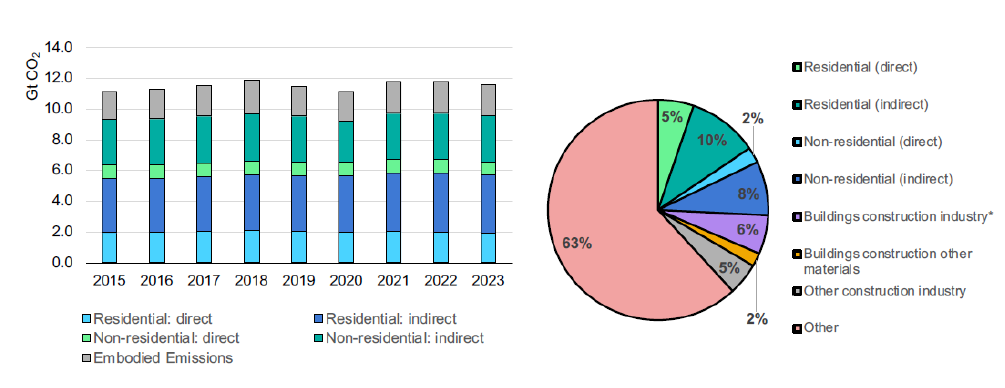}
	\caption{$\text{CO}_2$ Emissions from the building sector from 2010 to 2023 (left) and the share of buildings in global energy- and process-related emissions in 2023 (right). Source: reproduced from~\cite{UNEPGlobalABC2024BeyondFoundations}.}
	\label{fig:intro_schema}
\end{figure}

Therefore, practical decarbonization solutions are required, and improving envelope and system efficiency alone is often insufficient. In this regard, a promising direction is given by the design of strategies aimed at explicitly increasing the utilization of on-site renewable generation, predominantly solar. However, photovoltaic (PV) output is intermittent and often poorly aligned with building energy demand, so surplus electricity frequently occurs when it cannot be consumed directly. 
To manage the intermittency of energy production from PV systems, two strategies are typically adopted: feeding the generated energy into the grid and using electrochemical storage systems. However, both solutions come with some drawbacks. Indeed, in the first case, under a scenario of full electrification of the built environment, a large-scale injection of electrical energy into the grid may lead to over-voltage issues, overloads, and congestion in distribution lines, particularly in the presence of old electrical networks \cite{DAMIANAKIS2025125000,GUPTA2021116504}. Consequently, an excessive reliance on the electrical grid may become critical for the stability and reliability of the system. Considering batteries, instead, it must be noted that they are technically effective but not fully sustainable. In fact, some studies show that a non-negligible share of the environmental footprint is related to battery manufacturing and end-of-life processing, with outcomes strongly dependent on assumptions regarding chemistry, lifetime, depth of discharge, and recycling pathways  \cite{DEHGHANISANIJ2019192,DASILVALIMA2021101286}. Moreover, although recycling technologies are advancing, they are still complex and energy-intensive, and their viability depends on logistics and recovered material value, hindering both sustainability and cost at scale~\cite{DEHGHANISANIJ2019192,en10111760,LAMNATOU2020134269}.

These limitations have motivated growing interest in battery-free alternatives. In this context, the most promising solutions aim at increasing renewable self-consumption by exploiting intrinsic storage within the building. This perspective naturally connects to the concept of building energy flexibility, defined as the capability of a building to modulate demand and/or generation in response to climate, occupants’ needs, and grid requirements~\cite{ALESCI2025135903,annex67}. Energy flexibility can be provided through dedicated storage (electrical or thermal) or, importantly, through passive storage enabled by the thermal capacity of envelope components~\cite{ALESCI2025135903}.
In this context, several studies highlight thermal inertia as a promising alternative to active storage. Specifically, these works analyze how to engineer the building envelope so that it can store and release heat, shifting Heating, Ventilation, and Air Conditioning (HVAC) operation and effectively acting as a “virtual battery”~\cite{ZHI2024109892,jossen2004operation}. However, much of the existing literature emphasizes envelope design and retrofit measures as a passive mean to enhance thermal storage capacity, rather than a tool that can be real-time managed by a proper control framework.

Beyond storage technologies, the strategy used to regulate the way in which they absorb and release energy must also be considered. In this regard, is indeed key to note that many building control strategies aim at minimizing energy use or electricity cost. However, these objectives are not generally equivalent to minimizing $\text{CO}_2$ emissions. In fact, as electricity-related emissions depend on the time-varying grid carbon intensity, the same energy saved at different times can lead to markedly different emission outcomes. Therefore, not only an alternative to battery storage is required but also a dedicated control strategy that explicitly account for carbon intensity, rather than treating energy or cost minimization as a proxy for decarbonization.

To address these issues, in this work, we propose an optimal battery-free control strategy to reduce carbon emissions by storing the surplus of solar energy in the building's thermal mass. This avoids battery-related environmental and end-of-life impacts and makes the solution easier to implement, also reducing costs.
Specifically, we design a control strategy that, considering the forecasts of real-time carbon emission intensity, PV generation, external temperature, and building energy consumption, computes the optimal fraction $(\alpha(k))$ of available surplus solar energy to be stored into the building's thermal mass. Thus, it shifts the indoor temperature setpoint upward in winter and downward in summer, while maintaining occupants’ comfort. In doing so, our strategy stores solar energy when it is available and reduces subsequent reliance on grid electricity, thereby lowering near-term $\text{CO}_2$ emissions. To evaluate the effectiveness of the proposed strategy, we consider three case studies in which we analyze the behavior of the room, but with different thermal mass, i.e., with low, medium, and high thermal capacity. For each case, we simulate its thermal dynamics with and without the proposed control strategy, enabling a direct comparison of its impact across different envelope characteristics.
In all the scenarios, the results show that our strategy yields considerable $\text{CO}_2$ emissions reduction, while maintaining occupants' comfort. Furthermore, as a side effect, it also allows for reducing the consumption of grid electricity, thus favoring cost savings.

The remainder of this paper is organized as follows: Section~\ref{sec:review} better details the state of the art on building energy management techniques and on the use of the building envelope for passive thermal storage. Then, Section~\ref{sec:method} details the proposed method and its implementation, also providing insights into the theoretical rationale that supports it. Next, Section~\ref{sec:models} describes the case study and the models used to evaluate our strategy, while Section~\ref{sec:res} presents and discusses the obtained results. Finally, Section~\ref{sec:concl} comments on the main contributions of the work and outlines its possible future directions.

\section{Related Work}
\label{sec:review}
This Section summarizes the state of the art in building energy management. To this end, we first review the most effective strategies proposed in the literature, then focus on approaches that use the building's thermal mass as passive storage. The goal is to outline their benefits and current limitations, to effectively position our work and highlight its novel contributions.

\subsection{Building energy management}
Building energy management strategies can be broadly grouped into three 
categories, each with distinct strengths and limitations.
A first research line focuses on demand side management. These approaches aims at reshaping intelligent device scheduling and user behavior through load shifting, peak reduction, and price-based operation. Representative examples include reinforcement learning-based schedulers for residential devices and aggregations~\cite{mocanu2018line, alfaverh2020demand}, game-theoretic or negotiation-based coordination schemes \cite{Chouikhi2019AGM}, and heuristic optimization, mainly employed in industrial applications \cite{electronics9010105}.
Despite promising results, demand side management strategies depend heavily on the availability of sufficiently flexible loads and on user acceptance, and their performance degrades when short-term demand forecasts are inaccurate.

A second family of approaches includes model-based control strategies, where a predictive model is used to optimize HVAC setpoints and operating variables over a receding horizon~\cite{elnour2022neural,en14071996}. Compared to occupancy- or air quality-driven modulation, these methods explicitly exploit forecasts of exogenous disturbances and operating conditions to anticipate system behavior and adjust the plant accordingly.
Their effectiveness, however, strongly depends on the availability of accurate predictive models. To this end, in many contributions, models are learned via machine- or deep-learning methods, which prove to be less computationally intensive, despite providing similar accuracy performance. However, they also come with drawbacks, e.g., changes in operating conditions or occupancy patterns may lead to incorrect predictions, thereby requiring frequent re-calibration \cite{ZHOU2024132636}. Conversely, physics-based models yield more generalizable predictions but lead to computationally demanding optimization problems that may be difficult to solve in real time on embedded controllers.

Similar limitations characterize the last category of literature approaches, which includes works based on deep reinforcement learning~\cite{gao2019energy, valladares2019energy}, as well as metaheuristics and swarm intelligence~\cite{wahid2019improved, kanna2025swarm}, or even broader deep-learning-enabled frameworks~\cite{selvaraj2023smart, arun2025investigating} for HVAC control and comfort enforcement. While these methods can handle 
complex, nonlinear dynamics and high-dimensional state spaces, they are typically sample-inefficient, sensitive to reward design, and difficult to deploy safely without extensive training. Moreover, the lack of 
interpretability limits user trust and regulatory acceptance, which are 
critical factors for real-world adoption.

All three categories share a common limitation: they are primarily designed to minimize energy consumption or cost, without explicitly accounting for the carbon intensity of the grid. New strategies are therefore required that go beyond conventional energy reduction by explicitly targeting decarbonization. This means that they should not be designed to use less energy, but to use energy at the right time, i.e., when the grid is cleaner and renewable availability is higher. Achieving this goal requires integrating real-time carbon intensity signals into the energy management decision-making process, which remains largely absent from the existing literature.
Beyond that, two additional limitations also emerge. First, practical deployability: many state-of-the-art approaches rely on deep learning or reinforcement learning, which are sensitive to changes in plant configuration and operating conditions, poorly interpretable, and computationally complex.
Therefore, more interpretable, lightweight solutions are needed to promote adoption and trust. 
Last, given the well-known limitations of electrochemical batteries in terms of cost, degradation, and environmental impact, there is a clear need for strategies that enable renewable surplus management through battery-free alternatives.

\subsection{Building thermal mass as a passive storage system}
Considering the latter limitation, i.e., the need for battery-free strategies, several solutions have been proposed in the literature.
One of the most promising research line uses the thermal mass of the building as a passive means to store thermal energy. Such a solution is effective, as it enables the absorption of heat when available in excess and its gradual release when needed, thereby stabilizing indoor temperatures, reducing peak energy demand, and enhancing building energy flexibility~\cite{HAN2025100224, HAN2024113834, 
	ALESCI2025135903}.

Such approaches can be categorized into three groups: sensible heat storage through a high-inertia envelope materials, which attenuate indoor temperature peaks~\cite{HAN2024113834, NAVARRO20161334}; latent heat storage via Phase Change Materials, which exploit phase transitions at nearly constant temperature, achieving storage capacities from 5 to 14 times higher than sensible materials~\cite{RATHORE2019723, SOARES201382, TRIPATHI2024110128}; and active exploitation of the building envelope as a thermal battery, integrated with photovoltaics and heating systems to support energy management and emission reduction~\cite{ZHI2024109892}.

In this work, we focus on the third approach. Specifically, we build upon existing literature in the field, which can be organized along three main objectives: cost reduction, peak demand reduction, and PV self-consumption maximization.
Considering cost reduction, a reference work in the field proposes a variable setpoint strategy --- $21.5^\circ C$ during low-price periods and $19^\circ C$ during high-price periods --- applied to a Nordic multi-apartment building. This strategy yielded annual cost savings of 5.2\% in 2015 and 7\% in 2022~\cite{Ramesh2023813}. A similar approach proposes a floating setpoint pre-cooling strategy to reduce electricity import costs, achieving a reduction of approximately 6.1\% through temporal load shifting~\cite{ZHOU2022204}.
Considering approaches aimed at reducing HVAC energy use, \cite{en19041035} shows that raising the cooling setpoint from $24^\circ C$ to $26-28^\circ C$ can reduce air-conditioning load by up to 40\% in the two hours following a demand response event. Furthermore, \cite{JOHRA2019115} shows that hysteresis-based setpoint control within a 0.5--4~K band reduces heat pump energy use by 28--41\% over a four-month heating period. Comparisons between lightweight and heavyweight buildings show that higher thermal mass reduces cooling demand by 67--75\% during heat waves~\cite{KUCZYNSKI2020116984}. Finally, \cite{DOMINKOVIC2018949} shows that, in district heating contexts, pre-heating strategies can render 5.5--7.7\% of total heating demand temporally flexible.
Finally, considering temperature modulation approaches aimed at increasing the self-consumption of locally produced energy, several benefits have been reported. In the presence of PV surplus, pre-heating or pre-cooling strategies with variable setpoints exploit the building thermal mass by shifting the indoor temperature toward the limits of the comfort range ($20-23^\circ C$ instead of a constant $21^\circ C$). Simulations show a reduction in electricity imported from the grid of 45--80\%, an increase in renewable self-consumption of 29--43\%, and a reduction in evening peak demand of 55--85\%, while maintaining acceptable comfort conditions~\cite{su12020553}. Similarly, in high-performance residential buildings equipped with PV systems and modulating heat pumps, a rule-based strategy modulating the indoor temperature setpoint by approximately $\pm2^\circ C$ relative to the nominal value reduces electricity purchased from the grid by up to 17\% and increases PV self-consumption by 22\%~\cite{en13236282}.

Despite the breadth of these contributions, two critical gaps persist in the literature.
As previously mentioned, none of these works explicitly targets $\text{CO}_2$ emission reduction as the primary objective of the designed strategies.
In fact, even when using thermal storage, the presented strategies optimize cost or energy consumption, implicitly assuming that reducing energy use is equivalent to reducing emissions. 
Second, model complexity, interpretability, and computational cost are rarely reported, making it unclear whether proposed strategies can be realistically adopted in day-to-day building operation.
Therefore, in this work, we propose an optimization strategy that exploits building thermal mass as a battery-free thermal storage support, and explicitly targets $\text{CO}_2$ emission reduction, while also accounting for occupants' comfort. To this end, unlike existing approaches, we explicitly consider the real-time grid carbon intensity signal, ensuring that energy is consumed when the grid is cleaner and not just in smaller amounts. Furthermore, we rely on interpretable and lightweight models to simulate the system thermal behavior and to perform the optimization, obtaining an approach that it suitable for practical, large-scale deployment.

\section{Methodology}
\label{sec:method}

As introduced in Section~\ref{sec:intro}, our approach is based on a simple yet novel idea: minimizing the emissions associated with building energy use by leveraging the building’s thermal mass as a passive storage resource. To this end, we design a control strategy that, at each time step $k$, based on the time-varying carbon intensity value of the National Grid, computes the optimal fraction $\alpha(k)\in[0,1]$ of the available renewable surplus which should be converted into thermal energy to be stored in the building mass. 
It is important to note that this approach requires no modifications to the existing temperature control system. Instead, the proposed strategy virtually increases or decreases the temperature setpoint for a specific interval to store the surplus energy. In this way, the strategy pre-charges the building's thermal mass, and thus reduces future grid demand, as the indoor temperature remains near the user-defined setpoint using the stored energy. In this framework, $\alpha$ serves as the decision variable to optimize the amount of energy stored while ensuring the room temperature does not deviate excessively from the user's actual comfort settings.

More in detail, to achieve this goal, at each time step $k$, we optimize a single scalar decision $\alpha(k)$ and update it in a receding-horizon fashion; forecasts over the next $m$ steps of building energy consumptions, solar energy production, indoor and external temperature, and grid carbon intensity are used only to compute $\alpha(k)$, which is then applied to the current control action.
As previously mentioned, $\alpha(k)$ is chosen to reduce near-term grid electricity consumption and the associated $\text{CO}_2$ emissions, while also accounting for occupants’ comfort. 
Please note that, a key contribution of this work is going beyond aggregate $\text{CO}_2$ metrics; in fact, we explicitly consider temporal dynamics by using time-varying carbon intensity, avoiding average conversion factors that, although widely used in the literature \cite{KHAN20181091}, can mask when emissions occur.

\begin{figure}[H]
	\centering    \includegraphics[width=\linewidth]{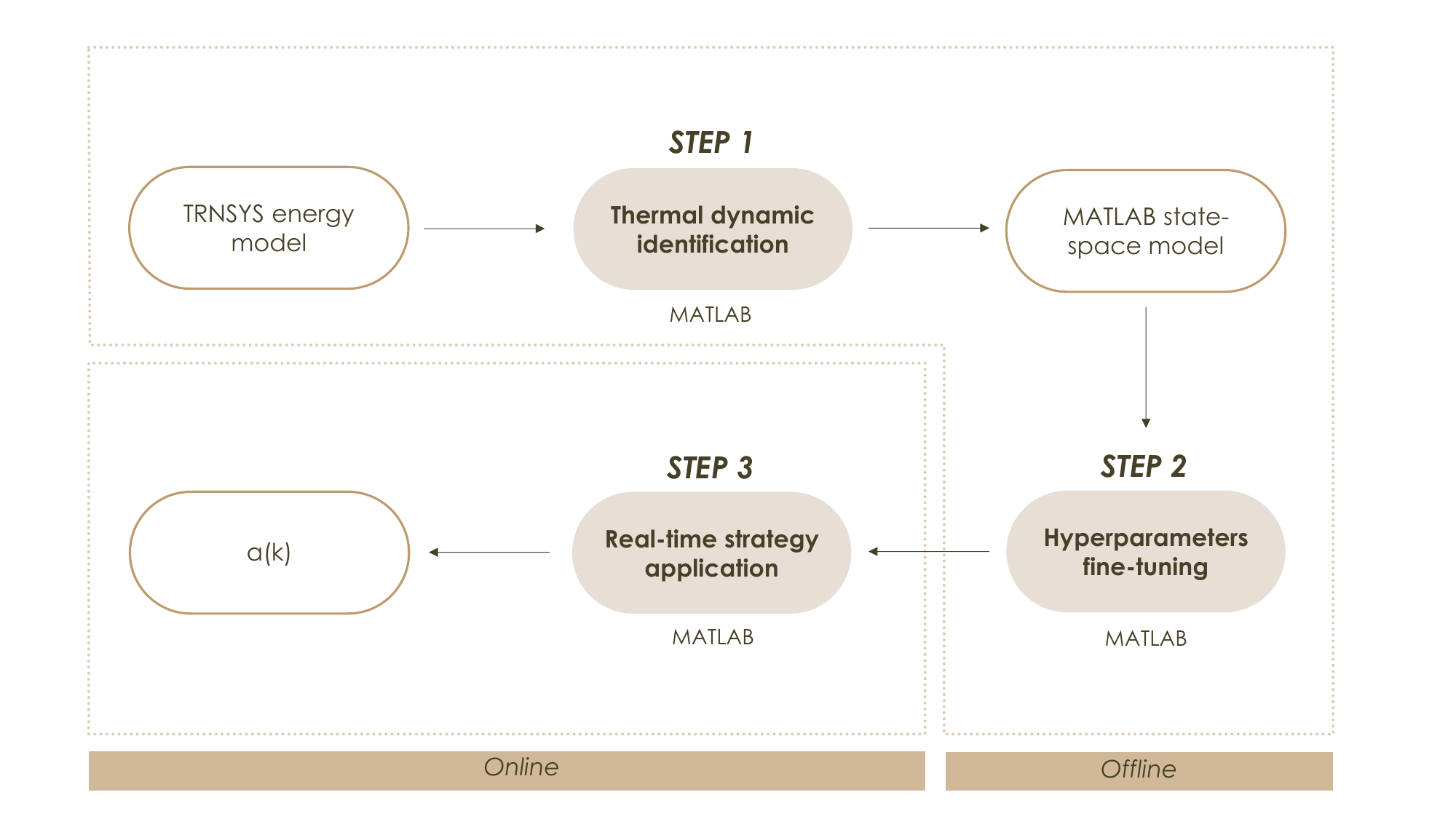}
	\caption{Flowchart illustrating the stepwise development, implementation, and application of the proposed optimization strategy.}
	\label{fig:Flowchart}
\end{figure}

Delving into the details of the approach, the steps followed for the development and implementation are outlined below. To facilitate a clearer understanding, a flowchart illustrating these steps is provided in Figure \ref{fig:Flowchart}.

It consists of 3 main steps, hereafter described:

\begin{itemize}
	\item STEP 1: thermal dynamic identification.The first step, described in detail in Section~\ref{sec:models}, consists of identifying an interpretable and simple model to describe the thermal behavior of the system of interest. This can be achieved by using data produced through detailed thermal simulations or collected from a physical building. In this case study, we use TRNSYS high-fidelity model to simulate the data required to identify a simplified state-space model of the system dynamics. Therefore, the TRNSYS model is used for initial identification and validation, while the reduced-order state-space model is the one actually used in the optimization strategy for real-time consumption forecasting.
	\item STEP 2: The energy consumption predicted by the reduced-order state-space model, combined with a forecast of solar energy production and time-varying grid carbon intensity data obtained from the Electricity Maps platform~\cite{electricitymaps}, is used to formulate the optimization problem and compute the optimal fraction $\alpha(k)$ of surplus solar energy to store in the building's thermal mass at each time step. The objective is to minimize emissions while enforcing comfort constraints. 
	As discussed later, the strategy depends on a set of hyperparameters that must be tuned for the specific system, namely the weight $\omega$, the sampling time $t_s$, and the prediction horizon $m$. This step is therefore devoted to the tuning of these hyperparameters via a Pareto-based procedure, which also clarifies how the chosen weights shape the comfort--emissions trade-off.
	\item STEP 3: real-time strategy application to the actual system. Once the optimal 
	hyperparameters are identified, the strategy can be used. Thus, given the identified state-space model, the number of occupants, solar energy generation forecast, and $\text{CO}_2$ intensity trends, the strategy determines how to adjust the temperature setpoint to reduce emissions by storing energy in the building's thermal mass when renewable generation is available, thereby reducing reliance on the grid at times of higher carbon intensity.
\end{itemize}
As reported in Section~\ref{sec:res}, in this work, we apply these steps to three different high-fidelity TRNSYS models. Such models are referred to the same room with varying thermal masses. Thus, for each case, we first build the TRNSYS model, and we then use it to generate the data required to identify the corresponding lightweight reduced-order model, used during optimization. Then, the optimal set of hyperparameters is computed for each room configuration, and the strategy is applied to a new scenario. For each case, the strategy results are compared against a baseline in which no thermal storage is performed, in terms of both $\text{CO}_2$ emissions reduction and occupants' comfort.

\subsection{Optimization strategy}
Delving into the details of the strategy implementation, $\alpha(k)$ is computed according to a receding-horizon procedure that accounts for the forecasts of building energy consumption ($E_{\mathrm{pred}}(k)$), the solar energy production ($E_{\mathrm{solar}}(k)$), and time-varying grid carbon intensity ($CI(k)$). Therefore, the objective function consists of two terms:
\begin{itemize}
	\item $\text{CO}_2$ emissions cost. To minimize this term, the control strategy should favor larger values of $\alpha(k)$ when the carbon intensity is expected to increase in the next hours, so that more solar energy surplus is stored and less grid energy is needed later.
	\item Thermal comfort. This term is introduced to penalize temperature deviations from the reference setpoint; therefore, $\alpha(k)$ is reduced when the induced temperature shift becomes too large.
\end{itemize}
Accordingly, the optimization problem is formulated as
\begin{equation}
	\label{eq:cost}
	J \;=\;
	\underbrace{\textcolor{DesignOrange}{J_{\text{CO}_2}}}_{\text{$CO_{2}$ emissions}}
	\;+\;
	\underbrace{\textcolor{DesignBlue}{J_{\mathrm{comfort}}}}_{\text{thermal comfort}} \, .
\end{equation}
In the following, we detail how $J_{\text{CO}_2}$ and $J_{\mathrm{comfort}}$ have been formulated.

\subsubsection{$\text{CO}_2$ Emission Cost}
This cost is introduced to reduce the energy drawn from the grid when the related carbon intensity is high. To this end, let $m$ be the length of the prediction horizon considered to perform the optimization. In the baseline case, where there is no surplus storage and the solar energy is only used instantaneously, the grid energy required over the horizon is:
\begin{equation}
	\label{eq:Egrid0}
	E_{\mathrm{grid},0}
	\;=\;
	\sum_{i=1}^{m}\max\!\big(E_{\mathrm{pred}}(i)-E_{\mathrm{solar}}(i),\,0\big),
\end{equation}
where $E_{\mathrm{pred}}(i)$ is the predicted building energy consumption at step $i$ and $E_{\mathrm{solar}}(i)$ is the available PV energy at the same step.

The corresponding baseline $\text{CO}_2$ emissions are then
\begin{equation}
	\label{eq:Jco2_0}
	J_{\text{CO}_2,0}
	\;=\;
	\sum_{i=1}^{m}\max\!\big(E_{\mathrm{pred}}(i)-E_{\mathrm{solar}}(i),\,0\big)\cdot CI(i),
\end{equation}
with $CI(i)$ denoting the time-varying grid carbon intensity signal.
Please note that, as previously mentioned, the choice of accounting for the temporal dynamics of the carbon intensity factor is a key contribution of this work, and is rarely addressed in the literature, which usually considers an average European factor.

The reported relations depend on $E_{\mathrm{pred}}(i)$, $E_{\mathrm{solar}}(i)$, and $CI(i)$. While the last two measures can be obtained from historical datasets (e.g., assuming cyclic behavior over a year) or forecasted using methods available in the literature \cite{8736879,10107594,SCOTT2023127807,ARYAI2023106314,KOHUT2025124527}, estimating $E_{\mathrm{pred}}(i)$ requires a model of the building thermal behavior in both heating and cooling operation, which can be identified from measured data or from physics-based simulations of the system. In this work we consider as a case study a room, whose complete behavior, as discussed in Section~\ref{sec:models}, has been modeled using TRNSYS software. However, as it was unnecessarily complex, we then model only the thermal behavior of interest using a state-space model, identified using ad-hoc simulated dataset from TRNSYS. This simplified model has then been used to predict the energy consumption of the room over time, which, along with solar energy and carbon intensity trends, have been used to implement our strategy.

Therfore, we are able to compute in real-time the actual emission cost, with our strategy implemented, which requires an additional term to be included with respect to the baseline formulation $J_{\text{CO}_2,0}$. Specifically, this term is related to the amount of energy saved thanks to the thermal storage. To derive it, we first compute the solar energy surplus as:
\begin{equation}
	\Delta E_{\mathrm{solar}}(i)
	\;=\;
	\max\!\big(E_{\mathrm{solar}}(i)-E_{\mathrm{pred}}(i),\,0\big).
\end{equation}
Then, we account that thanks to the control strategy, a fraction $\alpha(k)$ of this surplus is stored in the building thermal mass, providing an additional thermal energy input that can be computed as:
\begin{equation}
	\label{eq:delta_Q}
	\Delta Q(i)
	\;=\;
	\gamma\,\alpha(k)\,\Delta E_{\mathrm{solar}}(i),
\end{equation}
where $\gamma$ is the efficiency factor for the conversion from solar energy to thermal storage, which we set equal to $\gamma=1$ in the simulations.

To obtain a closed-form control law, we adopt the simplifying approximation that the stored thermal energy offsets future grid imports uniformly over the prediction horizon.
Accordingly, the actual $\text{CO}_2$-related cost can be computed as:
\begin{equation}
	\label{eq:Jco2}
	J_{\text{CO}_2}
	\;=\;
	\sum_{i=1}^{m}
	\max\!\left(
	\max\!\big(E_{\mathrm{pred}}(i)-E_{\mathrm{solar}}(i),\,0\big)
	-
	\frac{\alpha(k)}{m}\sum_{j=1}^{m}\Delta E_{\mathrm{solar}}(j),
	\,0\right)\, CI(i),
\end{equation}
where $\Delta E_{\mathrm{solar}}(i)$ and $\Delta Q(i)$ are defined as in the $\text{CO}_2$ cost formulation.

\subsubsection{Thermal Comfort Cost}
The thermal comfort cost, instead, is introduced to penalize any deviations of the indoor temperature from the reference setpoint. As a consequence, it limits the temperature change induced by storing solar energy surplus in the thermal mass. Starting from \eqref{eq:delta_Q}, the temperature variation due to the thermal storage at step $i$ can be approximated as:
\begin{equation}
	\Delta T(i) \;=\; \eta\,\frac{\Delta Q(i)}{C_{\mathrm{th}}}
	\;=\; \eta\,\frac{\gamma\,\alpha(k)\,\Delta E_{\mathrm{solar}}(i)}{C_{\mathrm{th}}},
\end{equation}
where $C_{\mathrm{th}}$ is the equivalent thermal capacitance and
\begin{equation}
	\eta \;=\;
	\begin{cases}
		+1, & \text{heating mode (temperature increase)},\\
		-1, & \text{cooling mode (temperature decrease)}.
	\end{cases}
\end{equation}

Consistently with the approximation used in \eqref{eq:Jco2}, the overall induced shift over $m$ steps is
\begin{equation}
	\Delta T_{\mathrm{tot}} \;=\; \eta\,\frac{\gamma\,\alpha(k)\sum_{i=1}^{m}\Delta E_{\mathrm{solar}}(i)}{C_{\mathrm{th}}}.
\end{equation}
Accordingly, we can define the comfort cost as:
\begin{equation}
	\label{eq:Jcomfort}
	J_{\mathrm{comfort}}
	\;=\; \big(\Delta T_{\mathrm{tot}}\big)^2
	\;=\; \alpha(k)^2\left(\frac{\gamma\sum_{i=1}^{m}\Delta E_{\mathrm{solar}}(i)}{C_{\mathrm{th}}}\right)^2.
\end{equation}

\subsubsection{Closed-Form Solution}
By combining \eqref{eq:Jco2} and \eqref{eq:Jcomfort}, we can rewrite \eqref{eq:cost} as:
\begin{equation}
	\label{eq:cost_extended}
	\begin{aligned}
		J =
		\textcolor{DesignOrange}{
			\left(E_{\mathrm{grid},0}-\frac{\alpha(k)}{m}\sum_{i=1}^{m}\Delta E_{\mathrm{solar}}(i)\right)
			\sum_{i=1}^{m} CI(i)} & +\\
		+\textcolor{DesignBlue}{
			\omega\,\alpha(k)^2
			\left(\frac{\gamma\sum_{i=1}^{m}\Delta E_{\mathrm{solar}}(i)}{C_{\mathrm{th}}}\right)^2
		} \, ,
	\end{aligned}
\end{equation}
where $\omega>0$ is a weight introduced to balance emissions reduction and thermal comfort. It can be tuned via a Pareto analysis to explore the comfort--emissions trade-off, and then selected according to application requirements.

Since $J$ is a convex quadratic function of $\alpha(k)$, the minimizer is obtained by setting $\frac{\mathrm{d}J}{\mathrm{d}\alpha(k)}=0$:
\begin{equation}
	\label{eq:dJdalpha}
	\begin{aligned}
		\frac{\mathrm{d}J}{\mathrm{d}\alpha(k)} = 
		-\left(\frac{1}{m}\sum_{i=1}^{m}\Delta E_{\mathrm{solar}}(i)\right)\left(\sum_{i=1}^{m} CI(i)\right) & + \\
		+ 2\omega\,\alpha(k)\left(\frac{\gamma\sum_{i=1}^{m}\Delta E_{\mathrm{solar}}(i)}{C_{\mathrm{th}}}\right)^2.
	\end{aligned}
\end{equation}
Solving for $\alpha(k)$ leads to a closed-form solution:
\begin{equation}
	\label{eq:alpha_star}
	\alpha^{\star}(k)
	\;=\;
	\frac{C_{\mathrm{th}}^{2}}{2\omega\,m\,\gamma^{2}}\,
	\frac{\sum_{i=1}^{m} CI(i)}{\sum_{i=1}^{m}\Delta E_{\mathrm{solar}}(i)}.
\end{equation}
If $\sum_{i=1}^{m}\Delta E_{\mathrm{solar}}(i)=0$, we set $\alpha^{\star}(k)=0$.
Finally, the actuation is saturated to satisfy $\alpha(k)\in[0,1]$:
\begin{equation}
	\label{eq:alpha_sat}
	\alpha(k) \;=\; \min\!\big(1,\max(0,\alpha^{\star}(k))\big).
\end{equation}

Please note that achieving this closed-form solution is a key advantage of this formulation as this considerably reduces the computational complexity requried to get optimal $\alpha(k)$ value at each step.

\section{Application to the Case Study}
\label{sec:models}
This section describes the simple yet high-fidelity reference system, consisting of a small office room modeled in TRNSYS, which has been considered to evaluate the effectiveness of our approach.

Specifically, three versions of the room are designed, each with a different thermal capacity. This allows us to demonstrate that the proposed strategy is effective regardless of the specific thermal properties of the building, and can therefore be applied without requiring any structural modification. Each room model serves as the ground-truth representation of the thermal behavior of the corresponding configuration. Based on data collected in ad-hoc performed simulations, we then identify a predictive discrete-time state-space representation of the system, which captures only the specific thermal dynamics of interest for our control strategy. Accordingly, we obtain a simple and interpretable model that is accurate enough to predict the heating and cooling power, while avoiding the computational complexity of the original TRNSYS counterpart, which is impractical for the optimization routine (STEP 1 in Figure~\ref{fig:Flowchart}).
Please note that, although the chosen system is intentionally simple, the proposed workflow is general and can be applied to more complex buildings, including using real operational data. Indeed, the identification step can be performed on any system to obtain a reduced-order state-space model linking the relevant exogenous variables to the heating and cooling power required. This separation between data source and control model makes the strategy flexible and system-independent.

Back to our considered case study, each room consists of an office located within a larger single-story building, situated in Milan. It has a net floor area of 30 $m^2$ (6 × 5 $m$) and a net internal height of 3 $m$. Three walls are adjacent to other spaces, whereas the remaining side is exposed outdoors. In Figure~\ref{fig:plant_building} of \ref{app:appendix}, the space representation in plan and section (along the AA' axis) is reported.

To investigate the effect of thermal mass on the strategy performance, three configurations are considered: light-, medium-, and heavy-weight mass. Their main characteristics are summarized in Table~\ref{tab:building_info}, which shows that all three configurations share the same thermal transmittance, but differ in their dynamic thermal properties, e.g., the decrement factor, the phase shift, the surface mass, and the internal areal heat capacity.  
In this study, particular attention is given to this last parameter, denoted as $C_m$, which represents the effective thermal capacity of each case study  \cite{iso13786_2008}. In fact, this is the main parameter determining the amplitude of the setpoint variation resulting from storing a fraction of the solar energy surplus in the thermal mass, as a function of the parameter $\alpha$.
Specifically, it has been computed as the sum of the internal areal heat capacities of the individual envelope components, which has been determined according to the dynamic thermal formulation provided by \cite{iso13786_2008}. According to~\cite{ASTE2015111}, it is calculated through the global heat transfer matrix of each building component, obtained by combining the transfer matrices of the individual layers across the construction.
To this end, we start from the thermal transfer equation:
\begin{equation}
	\begin{pmatrix}
		\hat{\theta}_2 \\
		\hat{q}_2
	\end{pmatrix}
	=
	\begin{pmatrix}
		Z_{11} & Z_{12} \\
		Z_{21} & Z_{22}
	\end{pmatrix}
	\cdot
	\begin{pmatrix}
		\hat{\theta}_1 \\
		\hat{q}_1
	\end{pmatrix}
\end{equation}
Here, $\hat{\theta}_1$ and $\hat{\theta}_2$ represent the complex amplitudes of temperature on the internal and external sides of the component, respectively, while $\hat{q}_1$ and $\hat{q}2$ denote the corresponding heat fluxes. The coefficients $Z{ij}$ are the elements of the thermal transfer matrix and describe the dynamic relationship between temperature and heat flux across the construction under periodic conditions.
From that equation, we then derive $\kappa_i$, which represents the internal areal heat capacity of the component over a considered period $T$, that is, by standard set to 24 hours:
\begin{equation}
	\kappa_i = \frac{T}{2\pi} \left| \frac{Z_{11} - 1}{Z_{12}} \right|
\end{equation}
Then, the internal heat capacity of the three case studies is determined by summing the internal heat capacities of all construction components, as defined before, each multiplied by its corresponding surface area:
\begin{equation}
	C_m = \sum_j \kappa_{ij} \cdot A_j.
\end{equation}

Back to the specific configuration, each room includes a glazed vertical element consisting of a window frame with double low-emissivity glazing, separated by a gas-filled argon cavity. Its solar heat gain coefficient (g-value) is 0.62, the visible light transmittance is 0.78, and the thermal transmittance is 1.1 $W/m^2K$.

\begin{table}[!h]
	\caption{Building envelope characteristics. This table reports the main properties of the building envelope based on the stratigraphy adopted for the design of the building hosting the considered office room, for each configuration analyzed in the case study. The stratigraphy labels are referred to Figures~\ref{fig:stratigraphy_light},\ref{fig:stratigraphy_medium},\ref{fig:stratigraphy_heavy} of \ref{app:appendix}.}
	\raggedright
	\resizebox{.99\textwidth}{!}{
		\begin{tabular}{c|cc|c|cccccc|c}
			\toprule
			\textbf{Building} & \multirow{2}{*}{\textbf{Classification}} & \multirow{2}{*}{\textbf{Configuration}} & \textbf{Stratigraphy} & 
			\textbf{Thickness} & \textbf{Thermal} & \textbf{Surface} & \textbf{Attenuation} & \textbf{Phase} & \textbf{Internal Areal Heat} & \textbf{Thermal}\\
			\textbf{Envelope} & & & \textbf{Subfigure} & \textbf{[$cm$]} & \textbf{Trans. [$W/m^2K$]} & \textbf{Mass [$kg/m^2$]} & & \textbf{Shift $[h]$} & \textbf{Capacity $[kJ/m^2K]$} & \textbf{Capacity $[kJ/K]$}\\
			\midrule
			\hline
			\multicolumn{11}{c}{\textbf{LIGHT-WEIGHT}}\\
			\hline
			\hline
			\textbf{External} & & & & & & & & & & \multirow{12}{*}{3130.83}\\
			\textbf{opaque} & Wall & External & (a) & 20.9 & 0.206 & 49 & 0.608 & 5.39 & 24.3 &\\
			\textbf{enclosure} & & & & & & & & & &\\
			\cline{1-10}
			\textbf{Internal} & & & & & & & & & &\\
			\textbf{vertical} & Wall & Boundary & (b) & 12.5 & 1.563 & 45 & 0.960 & 1.58 & 21.7 &\\
			\textbf{enclosure} & & & & & & & & & &\\
			\cline{1-10}
			\textbf{Horizontal} & & & & & & & & & &\\
			\textbf{flat} & Roof & External & (c) & 44.05 & 0.203 & 55 & 0.904 & 3.30 & 27.3 &\\
			\textbf{enclosure} & & & & & & & & & &\\
			\cline{1-10}
			\textbf{Horizontal} & & & & & & & & & &\\
			\textbf{enclosure} & Floor & Boundary & (d) & 58.2 & 0.235 & 327 & 0.236 & 11.46 & 31.8 &\\
			\textbf{against ground} & & & & & & & & & &\\
			\hline
			\hline
			\multicolumn{11}{c}{\textbf{MEDIUM-WEIGHT}}\\
			\hline
			\hline
			\textbf{External} & & & & & & & & & & \multirow{12}{*}{6531.77}\\
			\textbf{opaque} & Wall & External & (a) & 39 & 0.207 & 228 & 0.267 & 9.84 & 50.7 &\\
			\textbf{enclosure} & & & & & & & & & &\\
			\cline{1-10}
			\textbf{Internal} & & & & & & & & & &\\
			\textbf{vertical} & Wall & Boundary & (b) & 15 & 1.572 & 144 & 0.768 & 4.19 & 51.4 &\\
			\textbf{enclosure} & & & & & & & & & &\\
			\cline{1-10}
			\textbf{Horizontal} & & & & & & & & & &\\
			\textbf{flat} & Roof & External & (c) & 46.5 & 0.198 & 335 & 0.215 & 11.07 & 68.9 &\\
			\textbf{enclosure} & & & & & & & & & &\\
			\cline{1-10}
			\textbf{Horizontal} & & & & & & & & & &\\
			\textbf{enclosure} & Floor & Boundary & (d) & 37.9 & 0.260 & 559 & 0.120 & 12.15 & 44.7 &\\
			\textbf{against ground} & & & & & & & & & &\\
			\hline
			\hline
			\multicolumn{11}{c}{\textbf{HEAVY-WEIGHT}}\\
			\hline
			\hline
			\textbf{External} & & & & & & & & & &\\
			\textbf{opaque} & Wall & External & (a) & 44 & 0.210 & 428 & 0.087 & 13.49 & 59.4 & \multirow{12}{*}{8182.05}\\
			\textbf{enclosure} & & & & & & & & & &\\
			\cline{1-10}
			\textbf{Internal} & & & & & & & & & &\\
			\textbf{vertical} & Wall & Boundary & (b) & 21 & 1.547 & 306 & 0.380 & 7.97 & 66.5 &\\
			\textbf{enclosure} & & & & & & & & & &\\
			\cline{1-10}
			\textbf{Horizontal} & & & & & & & & & &\\
			\textbf{flat} & Roof & External & (c) & 44.5 & 0.204 & 597 & 0.146 & 11.36 & 92.4 &\\
			\textbf{enclosure} & & & & & & & & & &\\
			\cline{1-10}
			\textbf{Horizontal} & & & & & & & & & &\\
			\textbf{enclosure} & Floor & Boundary & (d) & 48.3 & 0.253 & 785 & 0.048 & 15.52 & 48.4 &\\
			\textbf{against ground} & & & & & & & & & &\\
			\hline
			\bottomrule
	\end{tabular}}
	\label{tab:building_info}
\end{table}

In each model, we consider a single thermal zone represented by an air node, corresponding to the volume of air assumed to have a uniform temperature. The adjacent rooms are assumed to share the same thermal conditions, while the outdoor-facing façade is oriented north and features three windows, each measuring 120 × 140 cm. The boundary conditions used in the modeling are summarized in Table~\ref{tab:boundary_conditions}.

\begin{table}[!h]
	\caption{Boundary conditions. This table reports the values used for the modeling of heating and cooling control, natural ventilation, infiltration, and internal gains, which are shared across the three implemented rooms.}
	\raggedright
	\resizebox{.99\textwidth}{!}{
		\begin{tabular}{c|c|c}
			\toprule
			\textbf{Parameter} & \textbf{Type} & \textbf{Value}\\
			\midrule
			\hline
			\multirow{4}{*}{Internal gain} & Artificial lighting & $6 W/m^2$, active during working hours (09:00–19:00)\\
			\cline{2-3}
			& \multirow{3}{*}{Occupancy} & $130 W$ per person (ASHRAE standard for moderately\\
			& & active office work [29]). During working days (09:00–19:00),\\
			& & the number of occupants varies between 0, 2, or 4; weekends: 0.\\ 
			\hline
			\multirow{4}{*}{Control and operation} & Hourly air change rate & \multirow{2}{*}{$0.6 ACH$}\\
			& (infiltration and natural ventilation) &\\
			\cline{2-3}
			& Heating setpoint (temperature) & Stepwise profile ranging between $20^{\circ}$ and $22^{\circ}$\\
			\cline{2-3}
			& Cooling setpoint (temperature) & Stepwise profile ranging between $25^{\circ}$ and $27^{\circ}$\\
			\hline
			\bottomrule
	\end{tabular}}
	\label{tab:boundary_conditions}
\end{table}

\begin{figure}[!h]
	\centering
	\includegraphics[width=\linewidth]{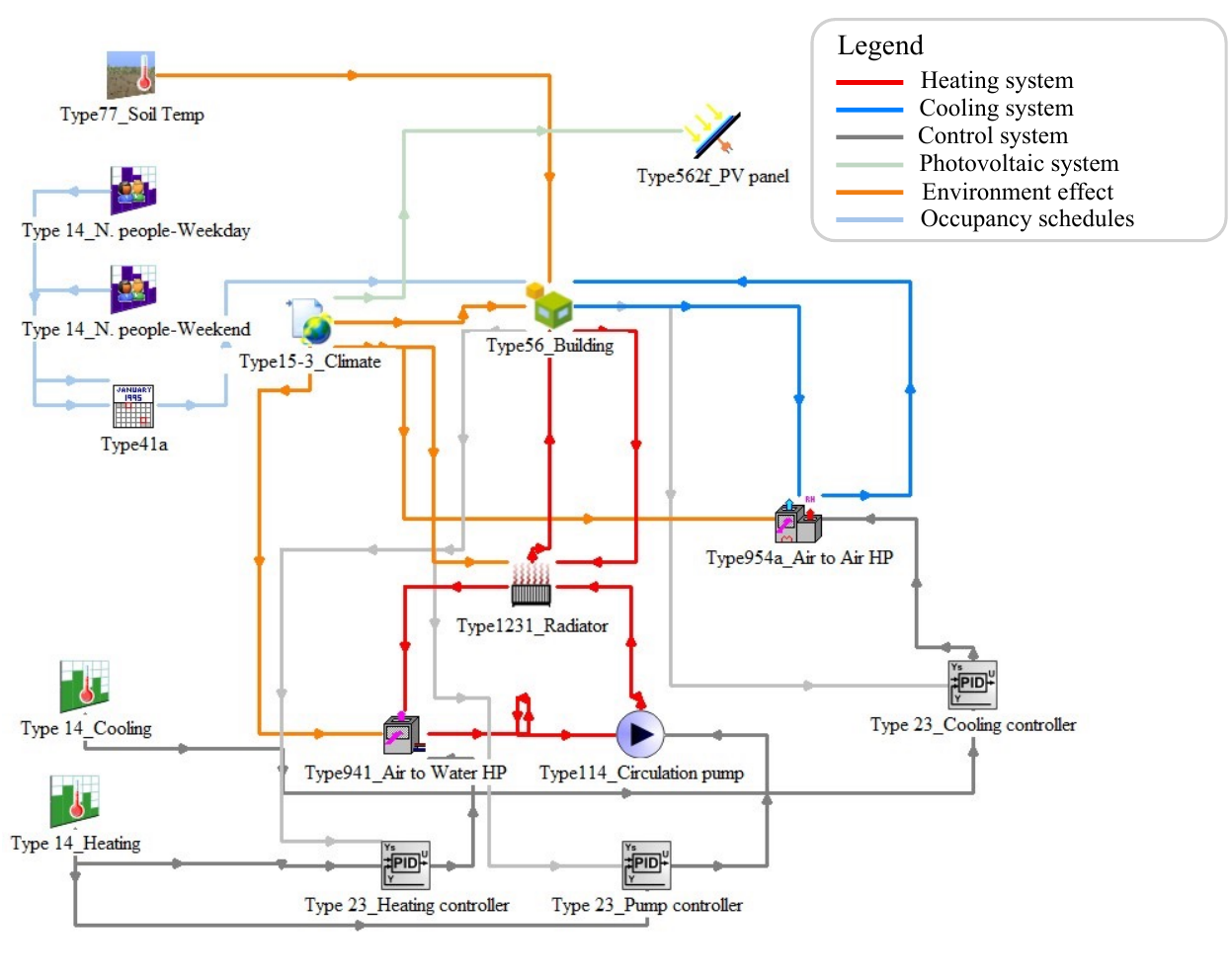}
	\caption{Plant of each TRNSYS model. This Figure provides an overview of the energy plant implemented in each TRNSYS room.}
	\label{fig:TRNSYS}
\end{figure}

Instead, Figure~\ref{fig:TRNSYS} shows the TRNSYS implementation of the energy plant implemented in each room, which will now be detailed. As reported, it consists of several blocks, defined Types, which correspond to a specific element of the plant and embed the respective governing equations. 

To provide a better overview of this plant, the composing types can be semantically grouped into six main circuits, represented using different colors. From a system perspective, the model represents a hybrid system for indoor climate control. Heating (red circuit) is provided by a high-temperature air-to-water heat pump (Type 941) for thermal generation and by a radiator (Type 1231) for heat distribution within the space. Cooling (blue circuit) is supplied by an air-to-air multi-split heat pump system (Type 954a), consisting of an outdoor unit connected to indoor unit.
Indoor comfort is maintained by a control system (gray circuit) based on two separate modulating PID controllers (Type 23). The controllers regulate the heating power of the air-to-water heat pump, the circulation pump power, and the cooling power of the multi-split system, based on the indoor air temperature and the reference setpoint. Effects of the thermal temperature and occupancy are instead described using the orange and light blue circuit, respectively.
Regarding on-site generation (green circuit), considering the single-store configuration of the building, a monocrystalline PV module with a nominal power of 445 $W$ is assumed. Accordingly, we design a system composed of 4 modules to provide a total peak power of 1.78 $kWp$. The PV system sizing was performed in accordance with Directive (EU) 2018/2001 (RED II) \cite{REDII} on renewable energy, meeting the requirements applicable to existing and new buildings.
Overall, in the sizing of the components considered, some parameters were customized to ensure consistency with the design conditions. The main configurations adopted for each Type are reported in Table~\ref{tab:TRNSYS_info} in the~\ref{app:appendix}.

\subsection{Thermal dynamic identification}
Once the TRNSYS models are implemented, the next step is to identify a simplified yet reliable surrogate that still characterizes the room's thermal response while being less computationally complex. This reduced-order model is used in the optimization process, while the TRNSYS counterpart is only used as a high-fidelity reference for model identification and validation.

It is worth noting that, since the approximated model is identified from data, whether simulated or collected from a real system, the same workflow can be applied to other buildings and to real operational measurements.
As discussed in Section~\ref{sec:intro}, existing literature generally usually perform the identification step in two different ways. On the one hand, data-driven approaches use regression predictors, e.g., gradient-boosted trees or recurrent neural networks, trained on environmental data, control variables, and occupancy proxies to forecast consumption. On the other hand, grey-box approaches model the room as a low-order RC (resistance-capacitance) network and estimate energy consumption by mapping thermal loads using simplified equipment models.
In this work, we leverage a reduced-order state-space representation model, as more interpretable and lightweight than the machine-learning counterpart. Specifically, we identify six models, two for each of the three TRNSYS models, one for heating and one for cooling behavior. Each model has been identified using the MATLAB \texttt{ssest} function~\cite{mathworks_sysid_doc} and discretized at the controller sampling time $ts$ for being integrated in the optimization framework. Although the designed TRNSYS models provide many signals, we select a reduced input set based on both correlation analysis results and practical measurability in real deployment. Therefore, the selected regressors are:
\begin{itemize}
	\item the setpoint temperature $T_{\mathrm{ref}}(k)$;
	\item the occupancy level $n_{\mathrm{occ}}(k)$ (used as an internal-gains proxy);
	\item the external temperature $T_{\mathrm{ext}}(k)$;
\end{itemize}
while the output variable is the related heating and cooling required power $P(k)$.
To ensure that the identified models capture the full range of thermal dynamics, for each TRNSYS room model we generated training datasets for two distinct seasonal periods: heating (October 15 – April 15) and cooling (April 15 – October 15). In each scenario, we specify the time-series trends for the room setpoint $T_{\mathrm{ref}}(k)$, occupancy count $n_{\mathrm{occ}}(k)$, external temperature $T_{\mathrm{ext}}(k)$ (based on Milan Brera meteorological data), and the resulting heating and cooling power required power $P(k)$.
Please note that, to identify the system dynamics accurately, it is essential to subject the model to large variations and diverse step conditions. Consequently, we applied varying setpoint profiles: in the winter scenario, a stepwise profile ranging from 20 $^\circ C$ to 22~$^\circ C$ was adopted, whereas in the summer scenario, the profile ranged from 24 $^\circ C$ to 26 $^\circ C$.
In both scenarios, the occupancy profile varies over the course of the day: no occupants are present during night-time (00:00–08:00 and 20:00–24:00), two occupants are present during 08:00–12:00 and 16:00–20:00, and four occupants are present from 12:00 to 16:00.
It is worth noting that, although high-excitation conditions were necessary for model identification, the resulting model is fully general and can predict the room's thermal behavior under arbitrary inputs. It is therefore used in the optimization strategy to forecast heating and cooling energy consumption across scenarios that differ from those used during identification.
Therefore, at the end of this identification process, we obtain a state-space model capable of accurately predicting the heating and cooling power of our room while avoiding the prohibitive computational overhead of TRNSYS, which would otherwise make real-time optimization infeasible.

To select the optimal order for the state-space models and verify their 
ability to accurately reproduce the TRNSYS reference behavior, models of 
order 1 through 3 are identified and compared using two metrics, mostly used in the literature:
\begin{itemize}
	\item the coefficient of determination ($R^2$),
	\begin{equation}
		R^2 = 1 - \frac{\sum_{i=1}^{N_t}\left(y_{\mathrm{val,true}}(i)-y_{\mathrm{val,pred}}(i)\right)^2}
		{\sum_{i=1}^{N_t}\left(y_{\mathrm{val,true}}(i)-\bar{y}_{\mathrm{val,true}}\right)^2},
		\qquad
		\bar{y}_{\mathrm{val,true}}=\frac{1}{N_t}\sum_{i=1}^{N_t} y_{\mathrm{val,true}}(i);
	\end{equation}
	
	\item the normalized mean absolute error ($nMAE$), normalized with respect to the maximum true output value,
	\begin{equation}
		\mathrm{nMAE} = 100 \cdot \frac{\frac{1}{N_t}\sum_{i=1}^{N_t}\left|y_{\mathrm{val,true}}(i)-y_{\mathrm{val,pred}}(i)\right|}
		{\max_{i}\left|y_{\mathrm{val,true}}(i)\right|}.
	\end{equation}
\end{itemize}

\begin{figure}[!h]
	\centering
	\begin{subfigure}[b]{0.45\textwidth}
		\centering
		\includegraphics[width=\linewidth]{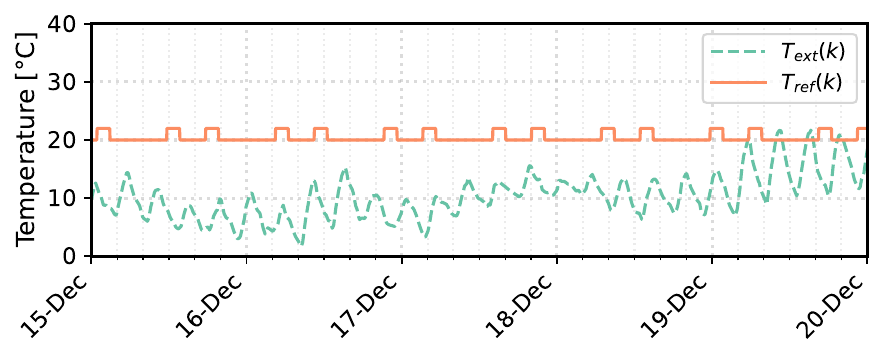}
		\caption{Winter: Temperatures ($T_{\mathrm{ref}}, T_{ext}$)}
	\end{subfigure}\hspace{0.7cm}
	\begin{subfigure}[b]{0.45\textwidth}
		\centering
		\includegraphics[width=\linewidth]{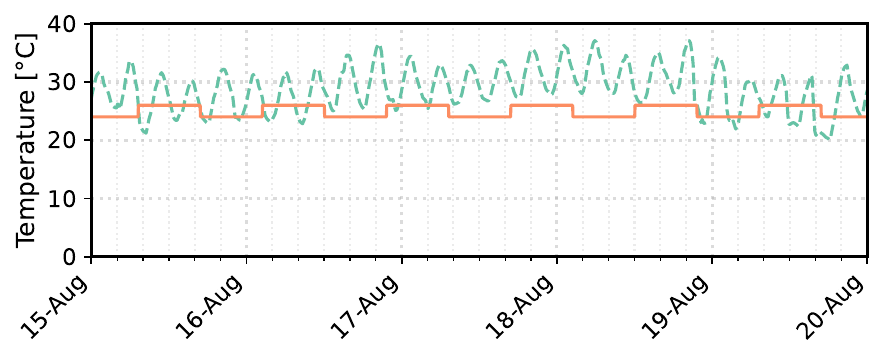}
		\caption{Summer: Temperatures ($T_{\mathrm{ref}}, T_{ext}$)}
	\end{subfigure}
	\begin{subfigure}[b]{0.45\textwidth}
		\centering
		\includegraphics[width=\linewidth]{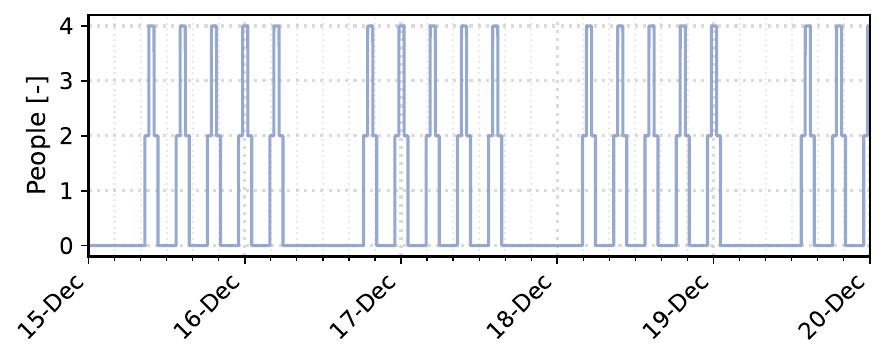}
		\caption{Winter: People ($n_{occ}$)}
	\end{subfigure}\hspace{0.7cm}
	\begin{subfigure}[b]{0.45\textwidth}
		\centering
		\includegraphics[width=\linewidth]{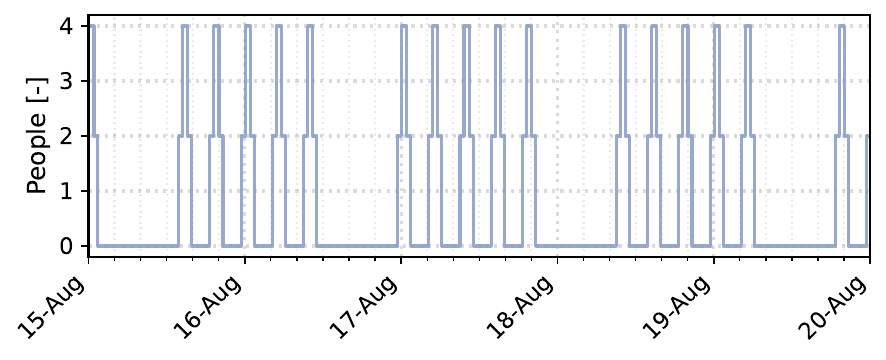}
		\caption{Summer: People ($n_{occ}$)}
	\end{subfigure}
	\begin{subfigure}[b]{0.45\textwidth}
		\includegraphics[width=\linewidth]{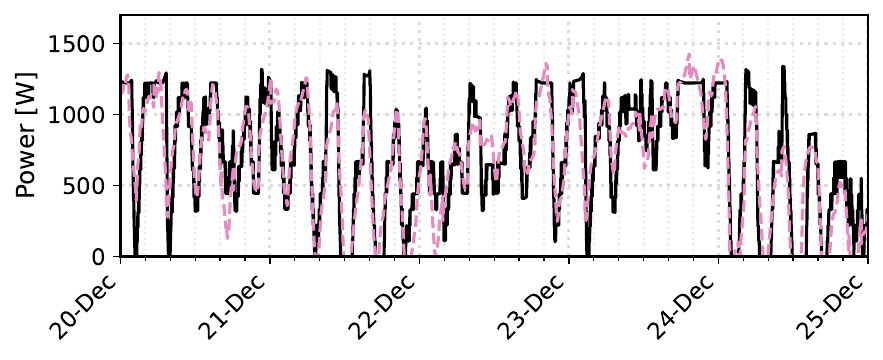}
		\caption{Winter: Heating Power ($P, P_{pred}$) - Light Configuration}
	\end{subfigure}\hspace{0.7cm}
	\begin{subfigure}[b]{0.45\textwidth}
		\includegraphics[width=\linewidth]{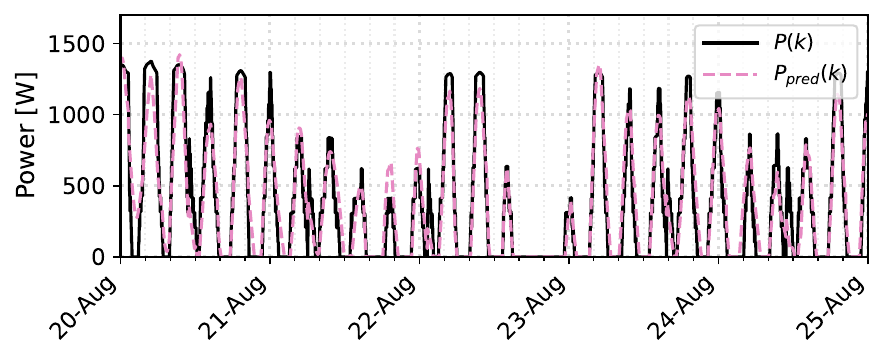}
		\caption{Summer: Cooling Power ($P, P_{pred}$) - Light Configuration}
	\end{subfigure}
	\begin{subfigure}[b]{0.45\textwidth}
		\includegraphics[width=\linewidth]{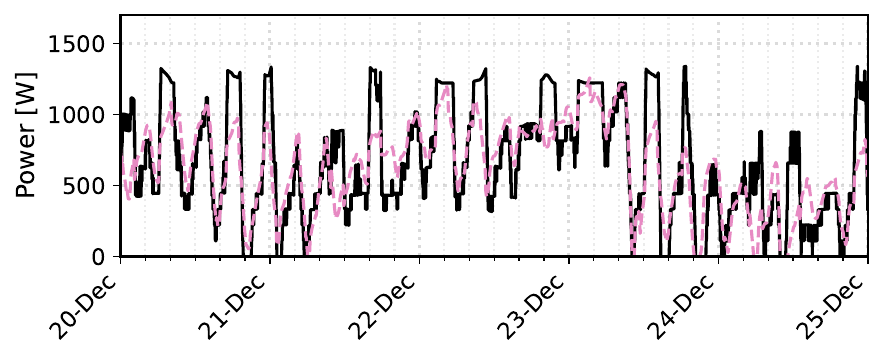}
		\caption{Winter: Heating Power ($P, P_{pred}$) - Medium Configuration}
	\end{subfigure}\hspace{0.7cm}
	\begin{subfigure}[b]{0.45\textwidth}
		\includegraphics[width=\linewidth]{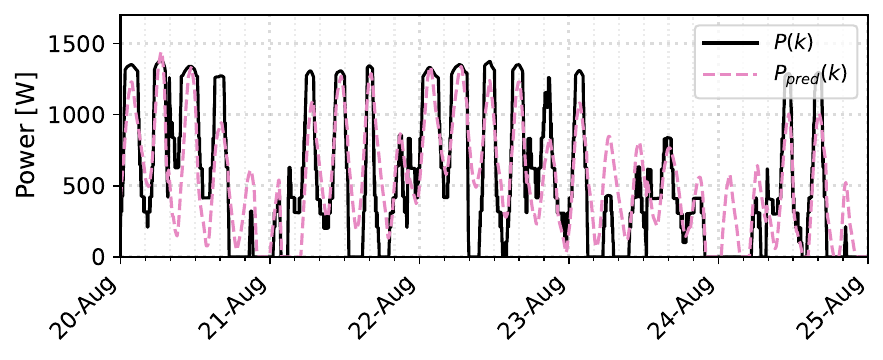}
		\caption{Summer: Cooling Power ($P, P_{pred}$) - Medium Configuration}
	\end{subfigure} 
	\begin{subfigure}[b]{0.45\textwidth}
		\includegraphics[width=\linewidth]{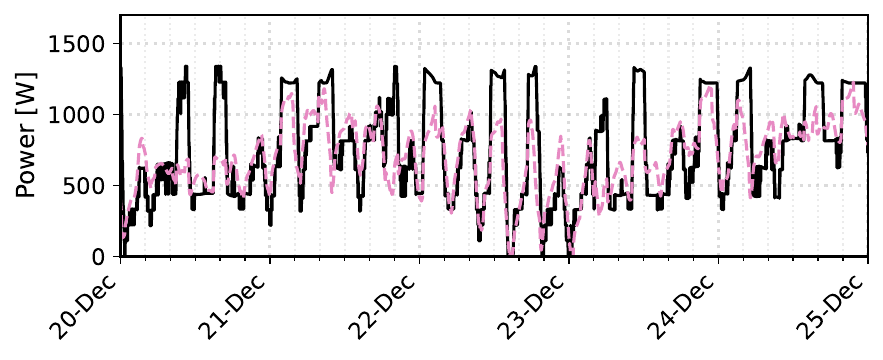}
		\caption{Winter: Heating Power ($P, P_{pred}$) - Heavy Configuration}
	\end{subfigure}\hspace{0.7cm}
	\begin{subfigure}[b]{0.45\textwidth}
		\includegraphics[width=\linewidth]{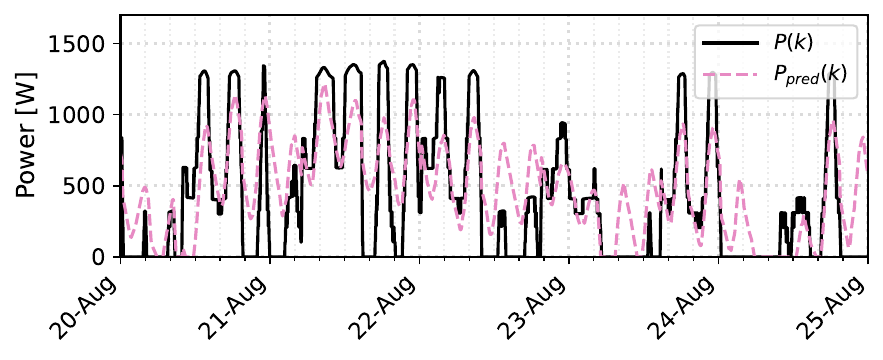}
		\caption{Summer: Cooling Power ($P, P_{pred}$) - Heavy Configuration}
	\end{subfigure} 
	\centering
	\caption{Heating and cooling model identification. This figure shows two representative validation periods, with winter data on the left and summer data on the right. The first two rows report the model's inputs, common to the three room configurations. The remaining rows compare the heating and cooling power predicted by the TRNSYS model ($P$) with that predicted by the corresponding surrogate model ($P_{\mathrm{pred}}$).}
	\label{fig:pred_vs_meas}
\end{figure}

In more detail, the evaluation is conducted on a dedicated TRNSYS-generated dataset, generated for each room in the conditions reported above and split into an identification and a validation subset. Each model is identified on the former and evaluated on the latter by computing $R^2$ and $nMAE$ considering the true TRNSYS output $y_{\mathrm{val,true}}(i)$ and the model prediction $y_{\mathrm{val,pred}}(i)$.
Results consistently show that a second- or third-order model provides the best accuracy across both heating and cooling conditions and for all three room configurations. First-order models fail to capture the dominant transient dynamics, while higher-order models yield only marginal accuracy improvements at the cost of increased complexity and reduced interpretability.
The aggregated results obtained when evaluating the best models identified for each room are reported in Table~\ref{tab:identification_res}, while a 
representative comparison between the temporal trends of TRNSYS and surrogate models predictions over 5 days of validation data is shown in Figure~\ref{fig:pred_vs_meas}. Overall, the identified models reproduces the reference thermal dynamics with good accuracy, thus supporting its use as the predictive model in the optimization procedure. The figure also highlights specific aspects of the room's thermal behavior that will be relevant when analyzing the results of the proposed strategy. In particular, the required power is generally higher as the building thermal capacity decreases, since lighter structures are less able to store heat and therefore respond more rapidly to external temperature variations. Consequently, the light configuration exhibits the highest demand, while the heavy configuration, thanks to its greater thermal inertia, shows the lowest. This behavior is more evident in summer simulations, where solar gains and temperature fluctuations are more intense, but it also characterizes the winter results, although to a lesser extent.

\begin{table}[H]
	\caption{Identification performance. This table reports, for each room configuration, the $R^2$ and $nMAE$ metrics achieved on the validation set by the surrogate heating and cooling models identified from the TRNSYS simulation.}
	\centering
	\resizebox{.8\textwidth}{!}{
		\begin{tabular}{c|cc|cc|cc}
			\toprule
			\textbf{Surrogate} & \multicolumn{2}{c|}{\textbf{Light-weight}} & \multicolumn{2}{c|}{\textbf{Medium-weight }} & \multicolumn{2}{c}{\textbf{Heavy-weight}}\\
			\textbf{Model} &\textbf{$R^2$} & \textbf{$nMAE$} & \textbf{$R^2$} & \textbf{$nMAE$} & \textbf{$R^2$} & \textbf{$nMAE$}\\
			\midrule
			\hline
			Winter & 0.77 & 10.24 & 0.64 & 13.73 & 0.60 & 14.00\\
			Summer & 0.90 & 6.10 & 0.76 & 10.52 & 0.70 & 12.24\\
			\hline
			\bottomrule
	\end{tabular}}
	\label{tab:identification_res}
\end{table}

\FloatBarrier

\section{Results and Discussion}
\label{sec:res}
Now that we have identified an efficient surrogate model for each TRNSYS room that accurately predicts the energy consumption (STEP 1 in Figure~\ref{fig:Flowchart}), our strategy can be applied. In fact, other information, such as the time series of grid carbon intensity $CI(k)$ and available solar energy $E_{\mathrm{solar}}(k)$ can be retrieved from measured data.
Specifically, as we assumed that our office is located in Northern Italy, we considered hourly data on grid carbon intensity $CI(k)$ (kgCO$_2$/kWh) from the Electricity Maps platform \cite{electricitymaps}, referring to the year 2024. The outdoor temperature profile, instead, was derived from the climatic data recorded at the Milano Brera meteorological station.
Last, the solar energy production of the PV system was estimated through simulations carried out using the room model developed in TRNSYS with the same climatic data. The simulation, conducted over an annual time horizon, allows the calculation of the electrical energy generated by the PV system, assuming a constant panel efficiency. 
Please note that the choice of using 2024 data has been made as we assume comparable seasonal patterns across years. However, as later discussed in the limitations subsection, if this assumption is considered not reliable, the same framework can be applied using any forecasting method from the literature to predict these variables in real-time.

With this setup, the proposed strategy is evaluated on each of the three 
room configurations over a full-year simulation spanning January~1 to 
December~31, 2024. During the heating season, the indoor temperature setpoint was set to $20^\circ\mathrm{C}$ during working hours (08:00–19:00) and $18^\circ\mathrm{C}$ during off-hours and weekends. Similarly, in the cooling season, we assume a setpoint of $26^\circ\mathrm{C}$ during working hours (08:00–19:00) and $28^\circ\mathrm{C}$ during off-hours and weekends.
Considering the occupation, during working hours, two people were considered to be consistently present in the office. 

\subsection{Strategy hyperparameters fine-tuning}
Before applying our strategy, its hyperparameters must be properly tuned for each room. As detailed in Section~\ref{sec:method}, they include the weight $\omega$, balancing the $\mathrm{CO}_2$-related and the comfort objectives, and the forecasting parameters, namely the sampling time $ts$ and the prediction horizon $m$, which together determine the temporal resolution and lookahead window of the optimizer. To properly set their values, two approaches are possible: if prior knowledge or application-driven constraints are available, the parameters must be set accordingly. Alternatively, as in this case, a sensitivity analysis can be performed to identify the most suitable ones.
To this end, prediction horizons of 12, 18, 24, and 48 hours are considered; for each horizon, sampling times of 30, 60, 120, 180, and 240 minutes are evaluated. For each $(m, ts)$ pair, $\omega$ is swept over a logarithmically spaced range from $1$ to $1e^{15}$, and the full one-year scenario is simulated. For each value of $\omega$, the resulting $\text{CO}_2$ emissions and comfort deviation define a point on the Pareto frontier; the optimal $\omega$ for that pair is then selected as the one that maximizes emission reduction while keeping comfort within acceptable bounds. Finally, the $(m, ts)$ combination that best achieves this objective across the full simulation is selected.
The identified optimal hyperparameters are then used to simulate the strategy on each of the three room configurations. The results are evaluated with respect to a baseline scenario, in which the same behavior of the same room is simulated without any thermal storage, i.e., setting $\alpha$ equals 0. Therefore, in the baseline case, we expect that $\Delta T$ is also 0, as the energy supplied to the room is exactly that required to meet the setpoint, while any available solar surplus is not used. On the other hand, for the same reason, the $\text{CO}_2$ emissions are expected to be higher than those measured when applying our strategy.
Therefore, to evaluate our strategy, we measure two indicators:
\begin{itemize}
	\item $\text{CO}_2$ emissions reduction, measured as the average grams saved per day and the overall percentage savings over a year, both computed relative to a baseline in which no thermal storage is applied;
	\item Temperature setpoint deviation, measured as the average and maximum deviation from the user-defined setpoint, relative to the baseline scenario in which the setpoint is followed exactly as no thermal storage is applied.
\end{itemize}
The optimal hyperparameter settings for each of the three room configurations are reported in Table~\ref{tab:performance}. 

The results show that the optimal prediction horizon $m$ scales as the configuration goes from light to heavy. This outcome is physically intuitive; in fact, the heavy configuration has greater thermal inertia. Therefore, the heat stored or released propagates more slowly through the structure. As a result, the strategy requires a larger horizon to anticipate charging and discharging decisions. The optimal sampling time $ts$, instead, ranges between $30$ and $60$ minutes, while longer values decrease the performance. Also, the optimal weight $\omega$ consistently assessed around $10^6$, suggesting that the trade-off between emissions reduction and thermal comfort is governed by the problem scaling rather than the specific thermal properties of the room. 

Considering savings and temperature deviations measured in these optimal configurations, some considerations should be made. In terms of emissions reduction, the medium and heavy configurations achieve substantially larger savings than the light-weight one, with annual reductions of approximately 25\% compared to 10\%. This confirms that greater thermal mass provides more storage capacity, enabling the strategy to shift a larger share of energy demand toward periods of low-carbon solar generation. 
Accordingly, temperature deviations also increase in medium and heavy configurations. Nevertheless, even when considering the heavy configuration, the average daily deviation remains lower than $0.4^\circ C$, with a maximum of $1.2^\circ C$, confirming that the strategy achieves significant emissions reductions without compromising occupant comfort.

\begin{table}[!h]
	\caption{Optimal hyperparameters set. The table reports the values of prediction horizon $m$, sampling time $ts$, and weight $\omega$ identified via grid search, along with the $\text{CO}_2$ emission reductions and setpoint deviation.}
	\centering
	\resizebox{.99\textwidth}{!}{
		\begin{tabular}{c|ccc|cc|cc}
			\toprule
			\multirow{2}{*}{\textbf{Configuration}} & \multicolumn{3}{c|}{\textbf{Hyperparameters}} & \multicolumn{2}{c|}{\textbf{Emissions Reduction}} & \multicolumn{2}{c}{\textbf{Setpoint deviation}}\\
			& \textbf{$m$ [h]} & \textbf{$ts$ [min]} & \textbf{ $\omega$ [-]} & \textbf{avg per day [g/day]} & \textbf{tot per year [\%]} & \textbf{avg per day [$^\circ C$]} & \textbf{max in a day [$^\circ C$]}\\
			\midrule
			\hline
			Light-weight & 12 & 30 & $10^6$ & 19.67 & 9.88 & 0.1 & 0.5\\
			Medium-weight & 24 & 60 & $10^6$ & 52.30 & 25.37 & 0.3 & 0.9\\
			Heavy-weight & 48 & 30 & $10^6$ & 46.06 & 24.77 & 0.4 & 1.2\\
			\bottomrule
	\end{tabular}}
	\label{tab:performance}
\end{table}

\begin{figure}[ht!b]
	\centering
	\begin{subfigure}[b]{0.45\textwidth}
		\centering
		\includegraphics[width=\linewidth]{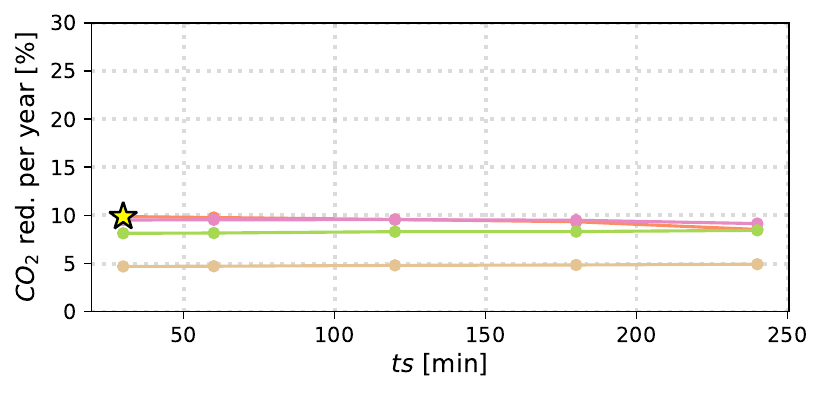}
		\caption{Average Emissions Reduction $[g/day]$ \newline - Light Configuration}
	\end{subfigure}
	\begin{subfigure}[b]{0.45\textwidth}
		\centering
		\includegraphics[width=\linewidth]{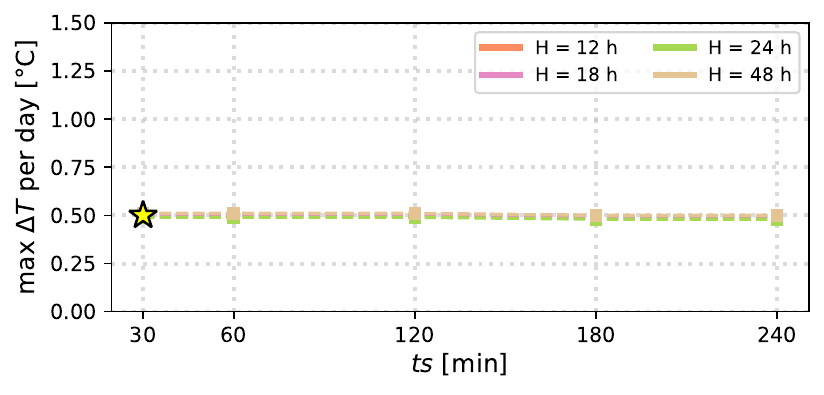}
		\caption{Average Setpoint Variation [$^\circ C$/day] \newline - Light Configuration}
	\end{subfigure}
	\begin{subfigure}[b]{0.45\textwidth}
		\centering
		\includegraphics[width=\linewidth]{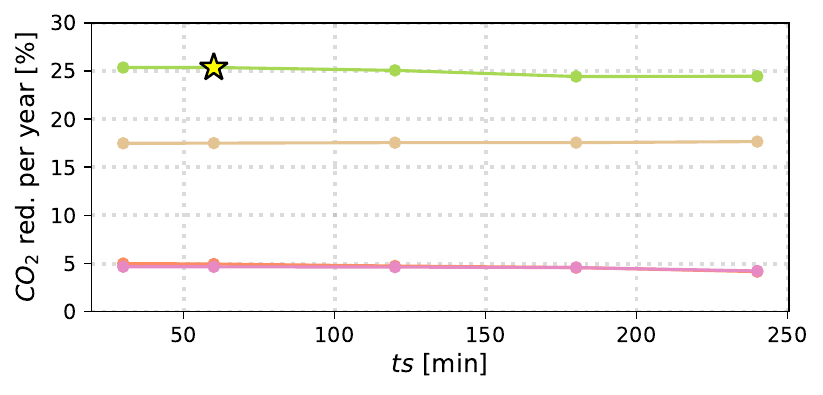}
		\caption{Average Emissions Reduction [g/day] \newline - Medium Configuration}
	\end{subfigure}
	\begin{subfigure}[b]{0.45\textwidth}
		\centering
		\includegraphics[width=\linewidth]{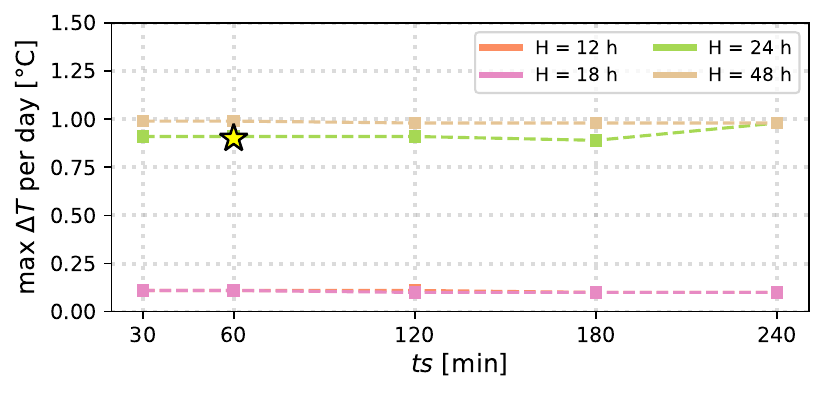}
		\caption{Average Setpoint Variation ($^\circ C$/day] \newline - Medium Configuration}
	\end{subfigure}
	\begin{subfigure}[b]{0.45\textwidth}
		\centering
		\includegraphics[width=\linewidth]{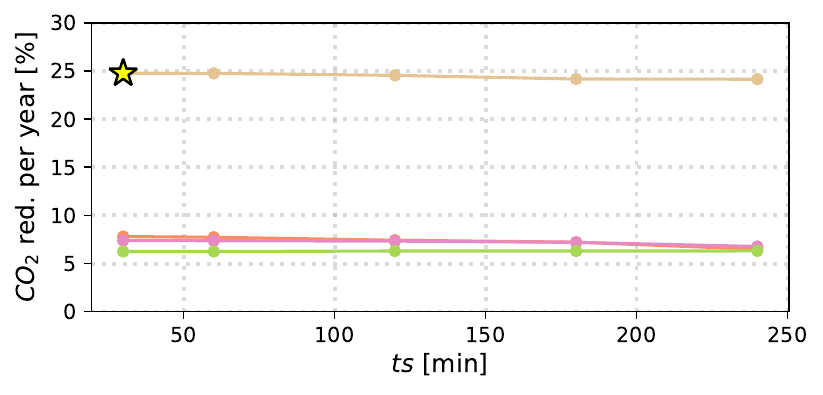}
		\caption{Average Emissions Reduction [g/day]\newline - Heavy Configuration}
	\end{subfigure}
	\begin{subfigure}[b]{0.45\textwidth}
		\centering
		\includegraphics[width=\linewidth]{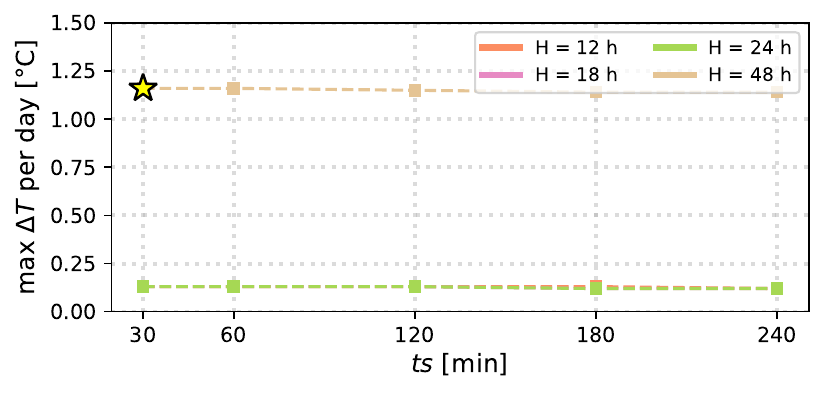}
		\caption{Average Setpoint Variation [$^\circ C$/day]\newline - Heavy Configuration}
	\end{subfigure}
	\caption{Sensitivity analysis on hyperparameters. This figure shows how total yearly emissions reduction (left) and maximum daily setpoint deviation (right) change as a function of the sampling time and the prediction horizon for each room configuration. Optimal hyperparameters set is marked by a yellow star.}
	\label{fig:performance}
\end{figure} 

Figure~\ref{fig:performance}, further extends the results in Table~\ref{tab:performance}, providing insights into how the yearly $\text{CO}_2$ reduction and the maximum daily $\Delta T$ varies as a function of the sampling time $ts$ and prediction horizon $m$ for the best-performing value of $\omega$ in the three different configurations analyzed.
These trends better highlight the intrinsic trade-off that characterizes our strategy: hyperprameteres set that allow for obtaining higher $\text{CO}_2$ emissions reductions also produce larger deviations from the temperature setpoint. Indeed, reducing the $\text{CO}_2$ cost requires increasing $\alpha$, which in turn increases the room temperature. As previously mentioned, this trend is more evident for the medium and heavy configurations, where the thermal inertia affects the heating and cooling dynamics of the system.
We also note that, in general, longer horizons (1 or 2 days) lead to greater $\text{CO}_2$ emissions reduction. This means that, as expected, higher look-aheads over surplus availability and carbon-intensity trajectories enable more effective temporal shifting of grid imports. 
Additionally, across all room configurations, higher actuation frequency, i.e., smaller $ts$, generally improves performance by allowing the controller to more effectively exploit the building's thermal mass.
Finally, regardless of the specific hyperparameters, the average daily deviation from the user temperature setpoint always remains below $\pm1.5^\circ C$, which is definitely within acceptable bounds to maintain a comfortable indoor environment.
Larger deviations are observed for the heavy model configuration, whereas lighter configurations reduce emissions without introducing significant temperature variations. 
Overall, this analysis shows that, although the optimal set of hyperparameters depends on the specific application requirements, a good general choice is to employ moderate-to-long prediction horizons, about 1 or 2 days, combined with a reduced sampling time, of approximately 30 minutes.

In the remainder of this section, more detailed results obtained by applying our strategy to each room configuration under the best hyperparameter settings, i.e., those reported in Table~\ref{tab:performance} will be presented.

\begin{figure}[ht!b]
	\centering
	\begin{subfigure}[b]{0.49\textwidth}
		\centering
		\includegraphics[width=\linewidth]{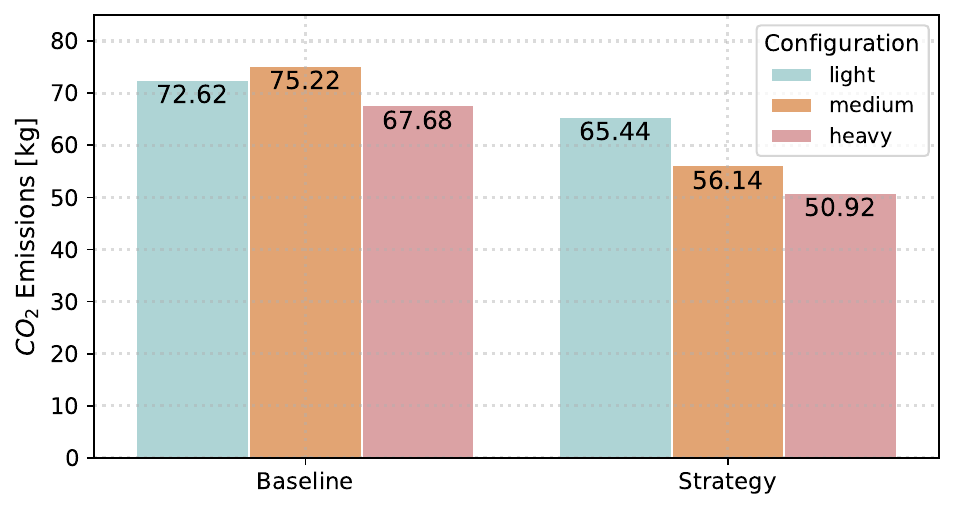}
		\caption{$\text{CO}_2$ Emissions [kg]}
	\end{subfigure}
	\begin{subfigure}[b]{0.49\textwidth}
		\centering
		\includegraphics[width=\linewidth]{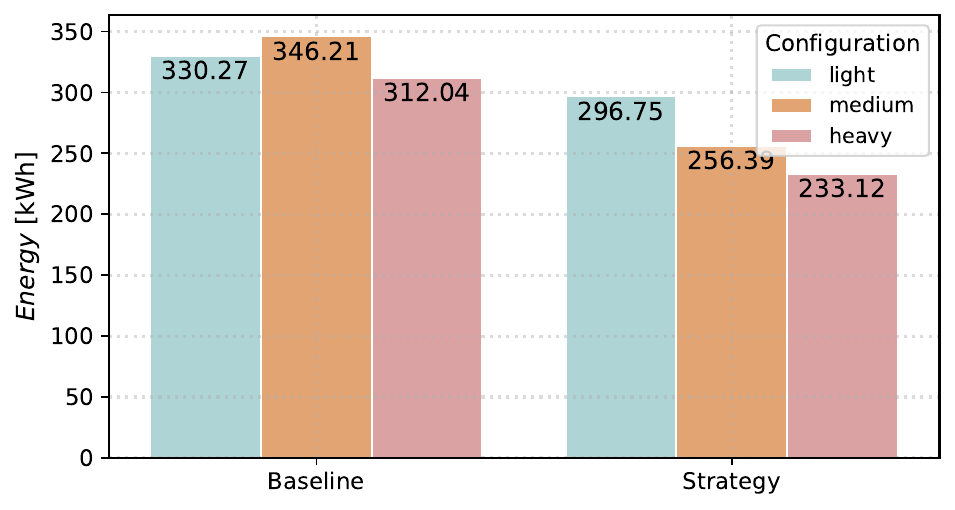}
		\caption{Energy Consumptions [kWh]}
	\end{subfigure}
	\caption{Strategy performance: annual results. This figure compares in all the considered room configurations the yearly aggregated $\text{CO}_2$ emissions (left) and energy consumption (right) for the baseline scenario ($\alpha(k)=0$) and the proposed strategy.}
	\label{fig:year_results}
\end{figure} 

\subsection{Strategy performance}
Once the optimal set of hyperparameters has been identified (STEP 2 in Figure~\ref{fig:Flowchart}), our strategy can be online applied (STEP 3) and its benefits can be investigated. To this end, a first comment can be made based on the results in Figure~\ref{fig:year_results}, which, for each room configuration, compares the annual $\text{CO}_2$ emissions and energy consumption provided by our strategy, with respect to the baseline scenario in which no thermal storage is applied. 
The results show that our strategy yields an annual $\text{CO}_2$ emissions saving with respect to the related baseline case of approximately $9.88\%$, $25.37\%,$ and $24.77\%$, for the light-, medium- and heavy-weight configurations respectively. This corresponds to $7.18$ kg, $19.09$ kg and $16.76$ kg of $\text{CO}_2$ saved in a year. In addition, although this is not the main objective of our strategy, we also highlight that it results in an annual energy saving with respect to the related baseline case of $10.15\%$, $25.94\%$, and $25.29\%$, corresponding to $33.52$ kWh, $89.81$ kWh, and $78.92$ kWh for the light-, medium-, and heavy-weight configurations respectively.

Beside $\text{CO}_2$ reduction, thermal comfort must also be considered. To this end, we used the Predicted Mean Vote (PMV) index, as defined by the UNI EN ISO 7730 standard \cite{fanger1970thermal, iso7730_2025} to evaluate how comfort varies when applying our strategy. Please note that, to compute this indicator, in accordance with ISO 7730:2005, metabolic rate and clothing insulation values were defined for both the summer and winter seasons, while additional parameters used for PMV calculation are reported in Table~\ref{tab:PMV_info} of~\ref{app:appendix}.
With this setup, the results demonstrate that our strategy achieves thermal comfort levels comparable to the baseline. Specifically, while the baseline PMV is $-0.31$ in winter and $0.33$ in summer, our approach yields similar values. This is further confirmed by the worst-case setpoint deviation, which remains limited to a PMV value of $-0.05$ in winter and $-0.04$ in summer.
Thus, we can conclude that PMV values fall within the limits defined by ISO 7730:2005, both for existing buildings (-0.7 < PMV < 0.7) and for new or renovated buildings (-0.5 < PMV < 0.5). Therefore, the proposed control strategy applies to both existing and new constructions, representing a valid tool to reduce emissions while meeting comfort constraints.

\begin{figure}[!t]
	\centering
	
	\hspace{0.2cm}\begin{subfigure}[t]{0.48\textwidth}
		\centering
		\includegraphics[width=\linewidth]{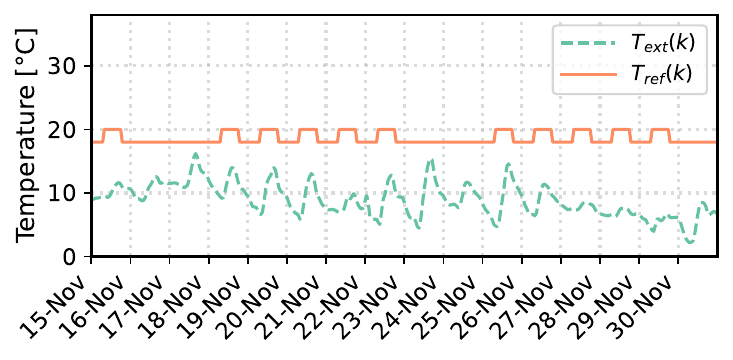}
		\caption{Winter: Temperatures ($T_{\mathrm{ref}}, T_{ext}$)}
	\end{subfigure}\hfill
	\begin{subfigure}[t]{0.48\textwidth}
		\centering
		\includegraphics[width=\linewidth]{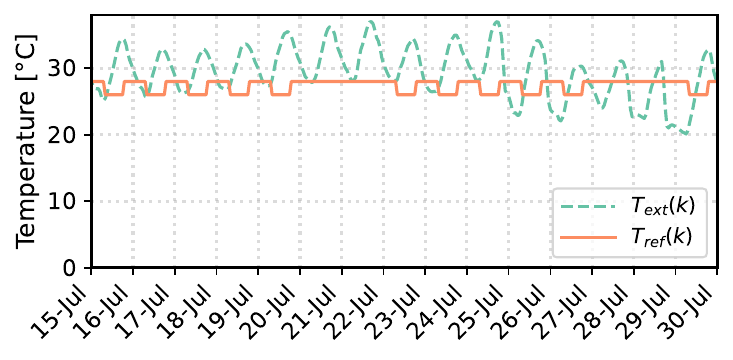}
		\caption{Summer: Temperatures ($T_{\mathrm{ref}}, T_{ext}$)}
	\end{subfigure}\\
	\hspace{0.2cm}\begin{subfigure}[t]{0.48\textwidth}
		\centering
		\includegraphics[width=\linewidth]{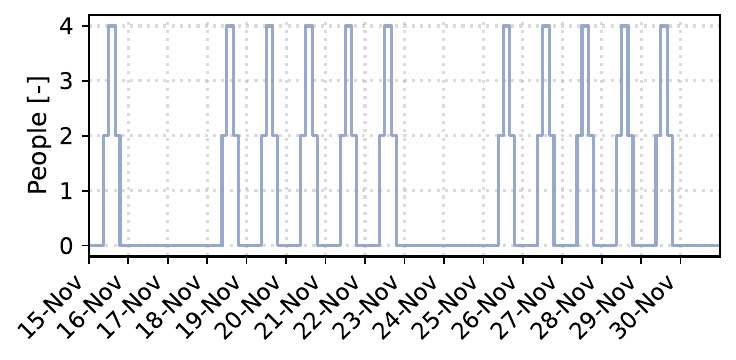}
		\caption{Winter: People ($n_{occ}$)}
	\end{subfigure}\hfill
	\begin{subfigure}[t]{0.48\textwidth}
		\centering
		\includegraphics[width=\linewidth]{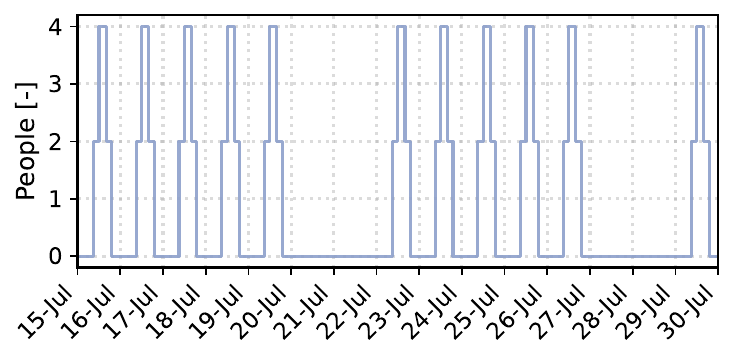}
		\caption{Summer: People ($n_{occ}$)}
	\end{subfigure}
	\begin{subfigure}[t]{0.5\textwidth}
		\centering
		\includegraphics[width=\linewidth]{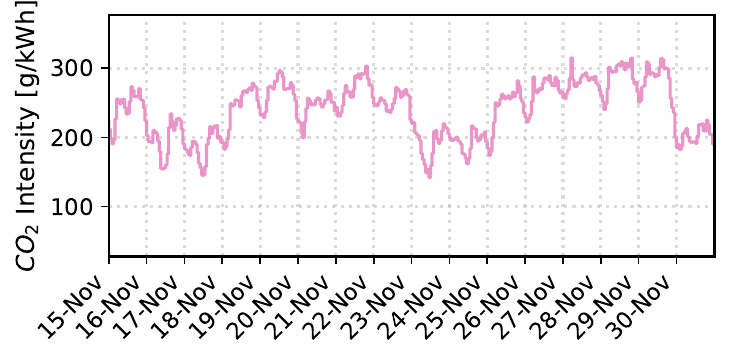}
		\caption{Winter: Grid Emissions Intensity ($CI$)}
	\end{subfigure}\hfill
	\begin{subfigure}[t]{0.5\textwidth}
		\centering
		\includegraphics[width=\linewidth]{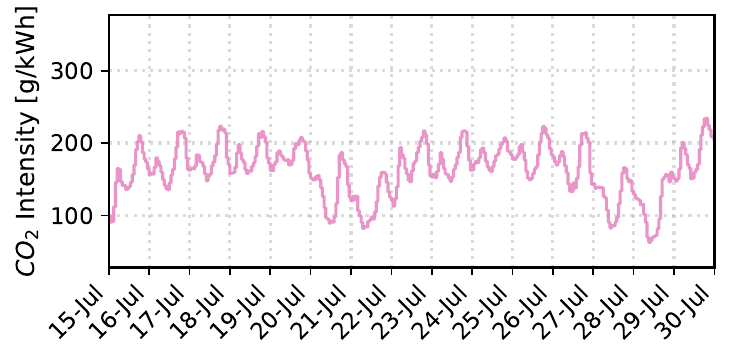}
		\caption{Summer: Grid Emissions Intensity ($CI$)}
	\end{subfigure}
	\caption{Simulated scenario: input parameters. This figure shows the trend of the input variables in a zoomed window of the defined scenario.}
	\label{fig:MPC_scenario_input}
\end{figure}

\begin{figure}[!t]
	\centering
	\begin{subfigure}[t]{0.48\textwidth}
		\centering
		\includegraphics[width=\linewidth]{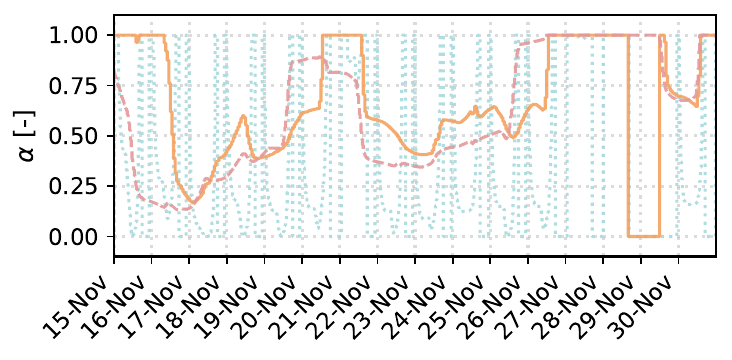}
		\caption{Winter: Surplus Fraction to Storage ($\alpha$)}
	\end{subfigure}\hfill
	\begin{subfigure}[t]{0.48\textwidth}
		\centering
		\includegraphics[width=\linewidth]{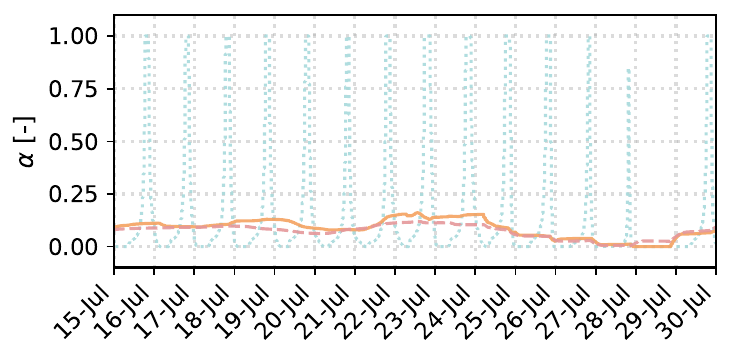}
		\caption{Summer: Surplus Fraction to Storage ($\alpha$)}
	\end{subfigure}
	\vspace{0.6em}
	\begin{subfigure}[t]{0.48\textwidth}
		\centering
		\includegraphics[width=\linewidth]{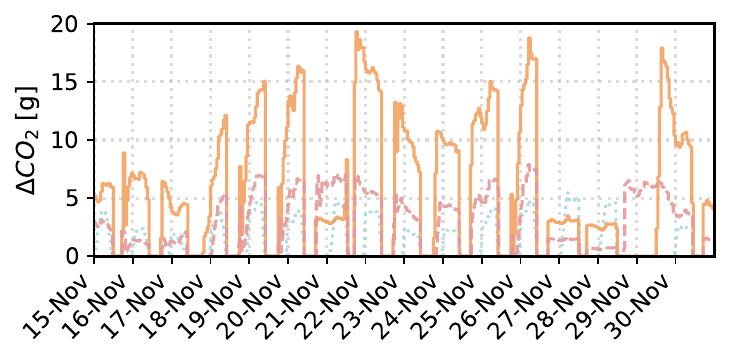}
		\caption{Winter: $\text{CO}_2$ Emission Reduction ($\Delta \text{CO}_2$)}
	\end{subfigure}\hfill
	\begin{subfigure}[t]{0.48\textwidth}
		\centering
		\includegraphics[width=\linewidth]{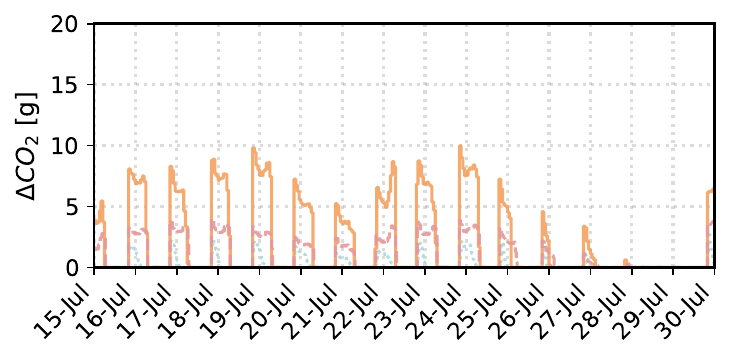}
		\caption{Summer: $\text{CO}_2$ Emission Reduction ($\Delta \text{CO}_2$)}
	\end{subfigure}
	\vspace{0.6em}
	\begin{subfigure}[t]{0.48\textwidth}
		\centering
		\includegraphics[width=\linewidth]{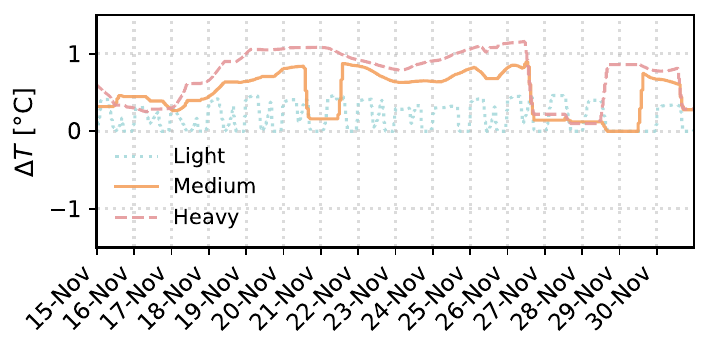}
		\caption{Winter: Temperature Variation ($\Delta T$)}
	\end{subfigure}\hfill
	\begin{subfigure}[t]{0.48\textwidth}
		\centering
		\includegraphics[width=\linewidth]{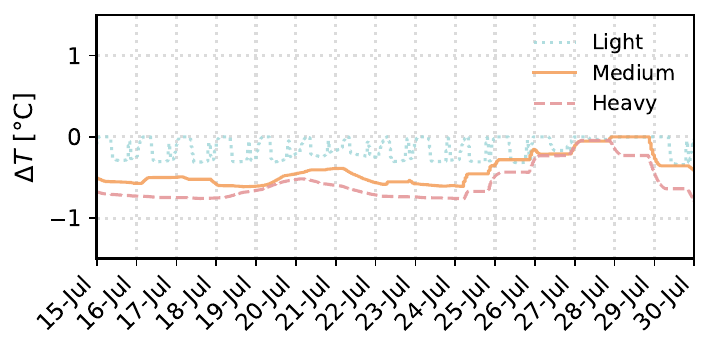}
		\caption{Summer: Temperature Variation ($\Delta T$)}
	\end{subfigure}
	
	\caption{Simulated scenario: output variables. This figure shows the trend of the output variables in the defined scenario, referred to a zoomed window of two weeks. Please note that $\Delta \text{CO}_2$ and $\Delta T$ are computed with respect to the baseline scenario in which no strategy is applied, i.e., ($\alpha=0$).}
	\label{fig:MPC_scenario_results}
\end{figure}

To better detail the outcomes of our strategy, Figures~\ref{fig:MPC_scenario_input} and~\ref{fig:MPC_scenario_results} show the trends of the relevant input and output variables defining our evaluation scenario over two representative zoomed windows. Those windows have been selected from the full-year simulation and span from the 15$^{th}$ to the 30$^{th}$ November for the winter period and from the 15$^{th}$ to the 30$^{th}$ July for the summer one. Please note that these specific windows have been reported as representative; the observed trends are, indeed, consistent across the rest of the year.

Specifically, Figure~\ref{fig:MPC_scenario_input} shows the input variables defining the scenario, i.e., the external temperature $T_{\mathrm{ext}}(k)$, the setpoint temperature of the room $T_{\mathrm{ref}}(k)$, the number of people $n_{occ}(k)$, the solar energy available $E_{\mathrm{solar}}(k)$, and the grid carbon intensity $CI(k)$. Figure~\ref{fig:MPC_scenario_results}, instead, shows the output of our strategy. In particular, in the first row, it reports the trends of the optimal surplus allocation factor $\alpha(k)$, i.e., the fraction of PV surplus that is suggested to store in the building thermal mass. In the middle row, instead, the trends of the related $\text{CO}_2$  emissions reductions are reported. These values have been computed, according to the surrogate emission model introduced in Section~\ref{sec:method}, as:
\begin{equation}
	\Delta \text{CO}_2(k) = \left(
	\max\!\big(E_{\mathrm{pred}}(k)-E_{\mathrm{solar}}(k),\,0\big)
	-
	\frac{\alpha(k)}{m}\sum_{i=1}^{m}\Delta E_{\mathrm{solar}}(i)
	\right) CI(k), 
\end{equation} 
i.e., as the difference of the grid-related emissions obtained in the baseline case, when $\alpha=0$, and those associated with the case in which the proposed strategy is applied.
Similarly, on the last row, the trends of the temperature variation are reported. These values have been computed as $\Delta T(k) = \eta\,\frac{\gamma\,\alpha(k)\,\Delta E_{\mathrm{solar}}(k)}{C_{\mathrm{th}}}$ (with $\gamma=1$ in our simulations), i.e., as the temperature deviation induced by the control action with respect to the baseline case ($\alpha=0$).

Analyzing these trends, we can derive some considerations. First, the limited thermal mass of the light-weight room hinders the capability of storing significant amounts of thermal energy. As a result, whenever surplus solar energy is available, our strategy sets $\alpha$ close to 1 in both winter and summer, thus storing nearly all excess power. However, due to the low thermal inertia, the room temperature rises rapidly, causing $\alpha$ to drop shortly after, as further storage would lead to high setpoint deviations. This behaviour produces an oscillatory temperature profile, which nonetheless remains within $\pm 0.5\,^\circ C$ of the setpoint, and yields small $\text{CO}_2$ savings. In contrast, the medium- and heavy-weight configurations exhibit a smoother $\alpha$ profile, reflecting the slower thermal response of the room. This also results in a more regular temperature evolution, but also in slightly larger deviations from the setpoint, which are, however, bounded within $\pm 1\,^\circ C$. Also, we can note that in winter, our strategy suggests storing most of the solar energy surplus across all configurations. In summer, instead, lower percentages of this energy are stored, especially considering the medium- and heavy-weight configuration. This is reasonable, as summer is characterized by larger solar surplus, so only a fraction of the available excess energy can be stored thermally to prevent excessive temperature increases.

\begin{figure}[t!]
	\centering
	\begin{subfigure}[b]{0.3\textwidth}
		\centering
		\includegraphics[width=\linewidth]{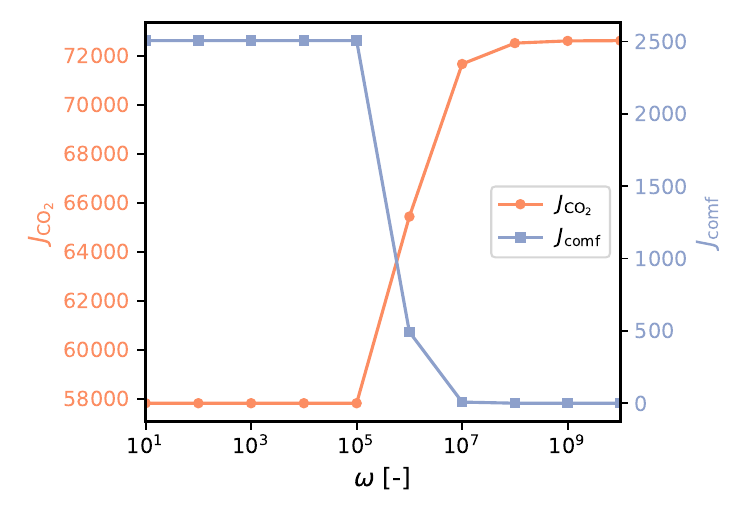}
		\caption{Light Configuration}
	\end{subfigure}
	\begin{subfigure}[b]{0.3\textwidth}
		\centering
		\includegraphics[width=\linewidth]{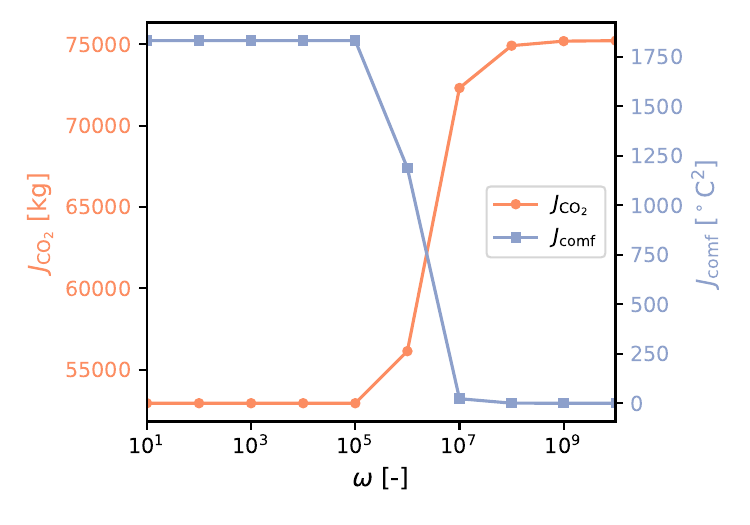}
		\caption{Medium Configuration}
	\end{subfigure}
	\begin{subfigure}[b]{0.3\textwidth}
		\centering
		\includegraphics[width=\linewidth]{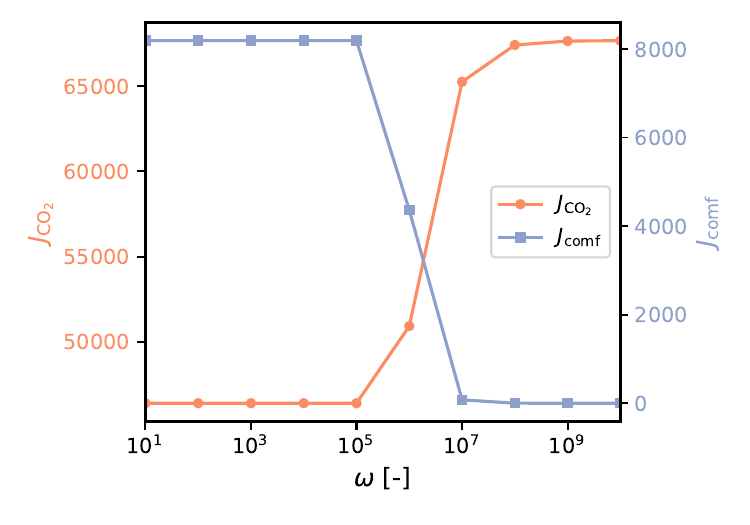}
		\caption{Heavy Configuration}
	\end{subfigure}
	\caption{$\omega$ fine-tuning. This figure shows the trends of annual $\text{CO}_2$ emissions reduction and average daily temperature deviation with respect to the weight $\omega$.}
	\label{fig:pareto}
\end{figure}

As a last comment, it is worth noting that the results discussed so far refer to the optimal configuration in Table~\ref{tab:performance}. Specifically, their refer to the optimal $\omega$ value, which provides the best balance between $\text{CO}_2$
emissions reduction and temperature setpoint deviation. However, depending on the specific application, a different trade-off may be desirable. To better understand the impact of this parameter on the strategy performance, Figure~\ref{fig:pareto} illustrates, for each room configuration, how both objectives vary as a function of $\omega$, keeping the prediction horizon and sampling time fixed to their optimal values. It is possible to note that, decreasing $\omega$ places greater emphasis on emissions reduction, allowing larger temperature deviations; conversely, increasing this parameter enforces adherence to the user-defined setpoint at the cost of reduced savings. This flexibility makes the proposed approach adaptable to a wide range of operational requirements, from comfort-critical environments to applications where emissions minimization is the primary objective.
\newline It is indeed possible to observe that, by forcing the weight toward a lower value, in order to induce a greater variation of the setpoint and increase the amount of solar surplus stored in the thermal mass, thus reducing emissions, an annual $\text{CO}_2$ saving of $27.29\%$, $30.04\%$, and $29.97\%$ can be achieved for the light-, medium-, and heavy-weight cases, respectively, compared to the baseline case.
These percentages correspond to reductions of 19.71 kg, 22.41 kg, and 20.43 kg of $\text{CO}_2$.
However, this increased saving also leads to a larger deviation of the setpoint and greater occupant discomfort, with maximum $\Delta T$ values of $2.6\,^\circ C$ $2.3\,^\circ C$, and $1.3\,^\circ C$, respectively. 

\FloatBarrier
\subsection{Limitations}
Despite the impressive results, it must be acknowledged that it is a proof-of-concept validation, and several simplifying assumptions have been adopted. Still, the scope of this work was to provide a preliminary insight into the potential of our control strategy to minimize $\text{CO}_2$ emissions through the exploitation of the building's energy flexibility provided by its thermal storage capacity. Therefore, it must be noted that none of the following limitations undermines this potential; rather, they point to clear extensions that can further improve robustness and applicability.

The first limitation is that the controller is evaluated using historical datasets to estimate solar energy production and external temperature, while a dataset from 2024 is used for the grid carbon intensity. In real deployments these signals must be forecast, and prediction errors may reduce the attainable savings. Nevertheless, the proposed control law only requires short-horizon trends and is computationally lightweight, making it straightforward to integrate standard forecasting pipelines.

Second, we assume that the energy offset induced by thermal storage is uniformly distributed over the prediction horizon. In other words, the amount of renewable surplus effectively shifted to the thermal mass is modeled as a constant reduction of grid energy across the next $m$ steps. This simplification is introduced to keep the problem analytically tractable and to preserve the closed-form solution for $\alpha(k)$, while still capturing the main effect of interest. In practice, the true impact of a setpoint shift on the energy trajectory may be non-uniform. Therefore, the uniform-allocation assumption can be relaxed in future work by considering other profiles for the surplus distribution. 

Third, the internal heat capacity is modeled as a 24-hour periodic signal, as required by the standard. This assumption is supported by the limited amplitude of its intra-day variations, which remain relatively small and repeat consistently over time. Therefore, the periodic approximation does not significantly affect the accuracy of the model and can be considered a reasonable simplification rather than a restrictive assumption.

Last, the proposed control strategy was validated in a simulated environment. Future work will involve its application to a real-world case.

Still, these limitations do not reduce the relevance of the approach; they rather indicate that the reported savings should be interpreted as a conservative lower bound obtained under simplified but realistic assumptions.

\section{Concluding Remarks}
\label{sec:concl}
In this work, we present a novel building decarbonization strategy based on a battery-free setup. Specifically, our idea is to store excess renewable energy, when available, in the building's thermal mass, using it as a controllable energy buffer.
To this end, we design an ad-hoc optimization strategy that, whenever a renewable surplus is available, computes the optimal fraction to store, trading off grid-related $\text{CO}_2$ emissions and occupants' comfort. Under specific assumptions, we show that a closed-form solution can be derived, yielding a policy that is lightweight and suitable for real-time implementation.

To assess the effectiveness of this approach, we consider a proof-of-concept validation by designing a simplified yet high-fidelity TRNSYS room model and identifying an interpretable discrete-time model for describing its thermal behavior, which has been then used for short-horizon energy consumption forecasting. The strategy is evaluated across three configurations of the case study, representing buildings with low, medium, and high thermal mass.
The results obtained by simulating our strategy performance over a year show a reduction in grid-related emissions of approximately $10\%$ compared to the reference scenario without storage in the low thermal mass case, corresponding to annual savings of $7$ kg of $\text{CO}_2$. In the medium and high thermal mass cases, the reduction reaches $25\%$ (approximatelly 18-20 kg of $\text{CO}_2$).  These savings are achieved with maximum setpoint deviations of $\pm 0.5^\circ C$ in the low thermal mass case, and $\pm 1.2^\circ C$ in the medium and high thermal mass cases. A thermal comfort assessment based on the PMV index confirms that, despite these deviations from the reference setpoint, thermal comfort remains within the limits prescribed by the standard.

Future work will focus on real-building validation, integrating forecasting models for solar energy and carbon intensity.
Overall, the presented results support the potential of the proposed intuition, showing that the building envelope as a thermal storage can be a low-cost alternative to batteries for increasing renewable self-consumption and reducing building-related $\text{CO}_2$ emissions.

\clearpage
% Appendix
\appendix
\section{Additional Details on TRNSYS Model}
\label{app:appendix}
\begin{figure}[!h]
	\centering
	\begin{subfigure}[b]{0.75\textwidth}
		\centering
		\includegraphics[width=\textwidth]{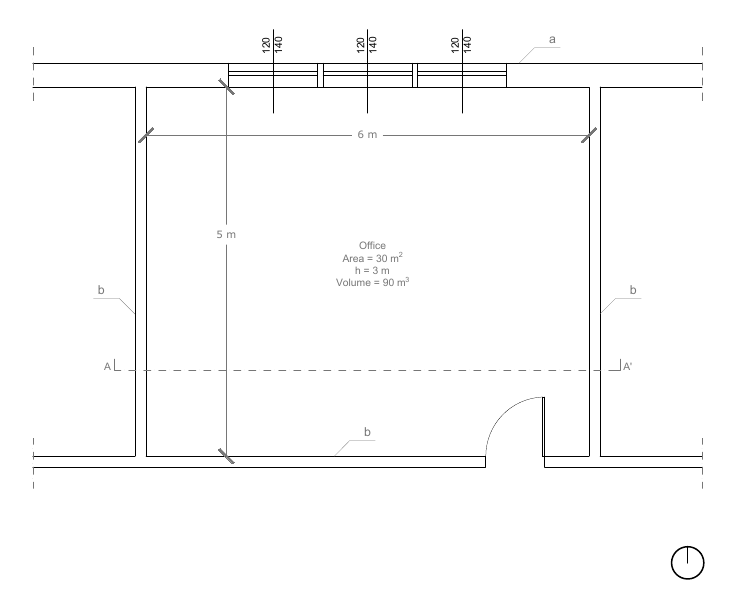}
		\caption{Plan}
	\end{subfigure}\\
	\begin{subfigure}[b]{0.75\textwidth}
		\centering
		\includegraphics[width=\textwidth]{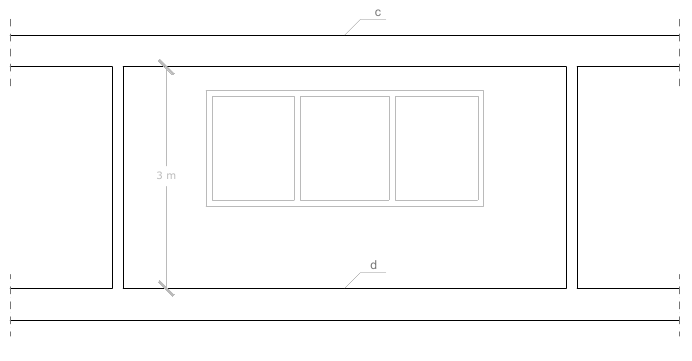}
		\caption{AA' Section}
	\end{subfigure}
	\caption{Designed building section. This Figure shows the plant and the AA' section of the building hosting the designed office room.}
	\label{fig:plant_building}
\end{figure}

\begin{figure}[!h]
	\centering
	\begin{subfigure}[b]{0.45\textwidth}
		\centering
		\includegraphics[width=\textwidth]{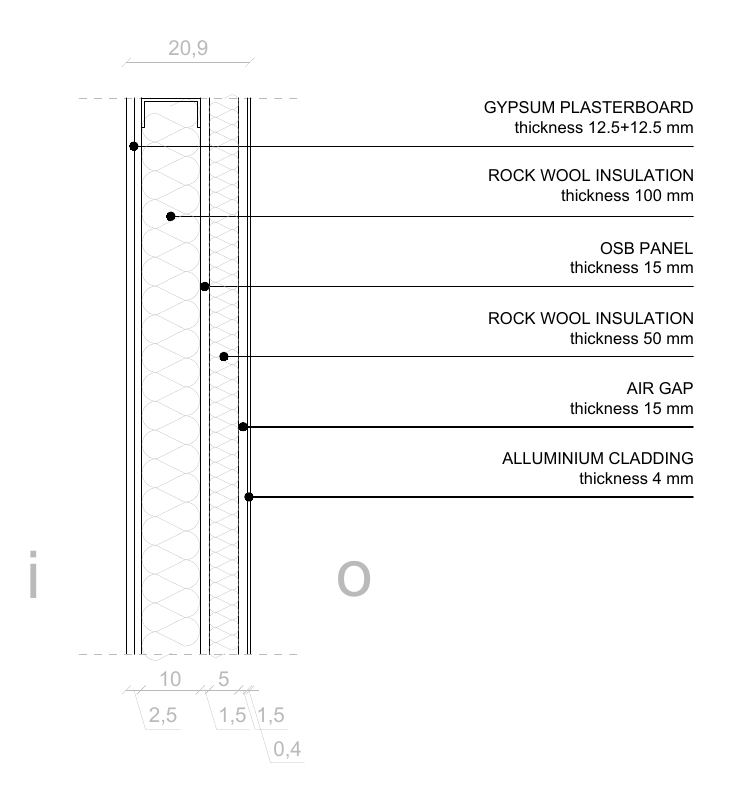}
		\caption{External opaque enclosure}
	\end{subfigure}
	\begin{subfigure}[b]{0.45\textwidth}
		\centering
		\includegraphics[width=\textwidth]{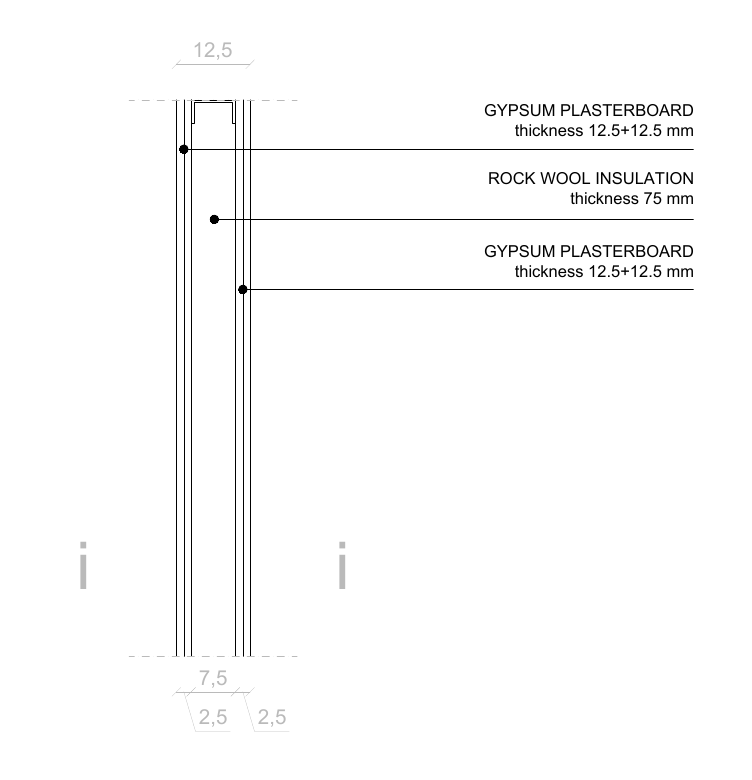}
		\caption{Internal vertical enclosure}
	\end{subfigure}
	\begin{subfigure}[b]{0.45\textwidth}
		\centering
		\includegraphics[width=\textwidth]{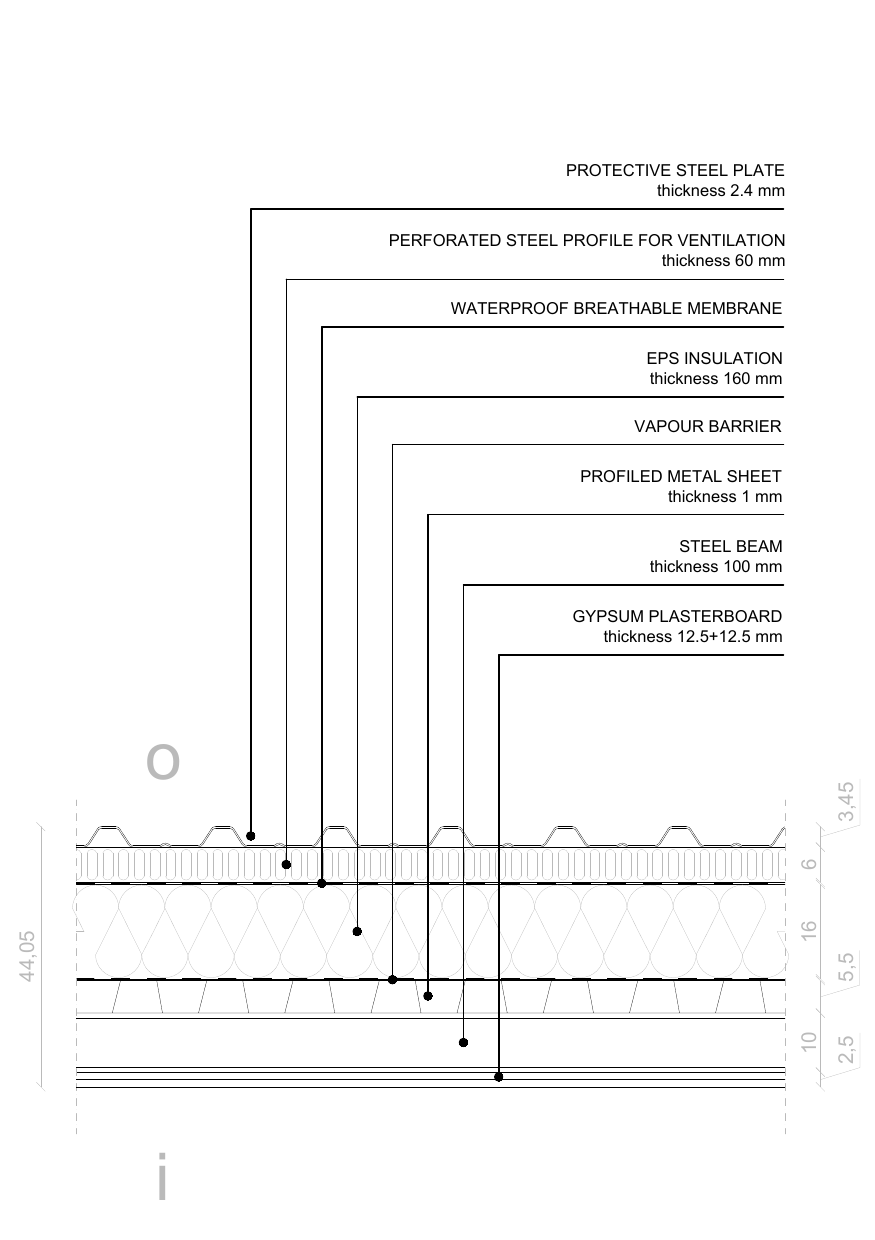}
		\caption{Horizontal flat enclosure}
	\end{subfigure}
	\begin{subfigure}[b]{0.45\textwidth}
		\centering
		\includegraphics[width=\textwidth]{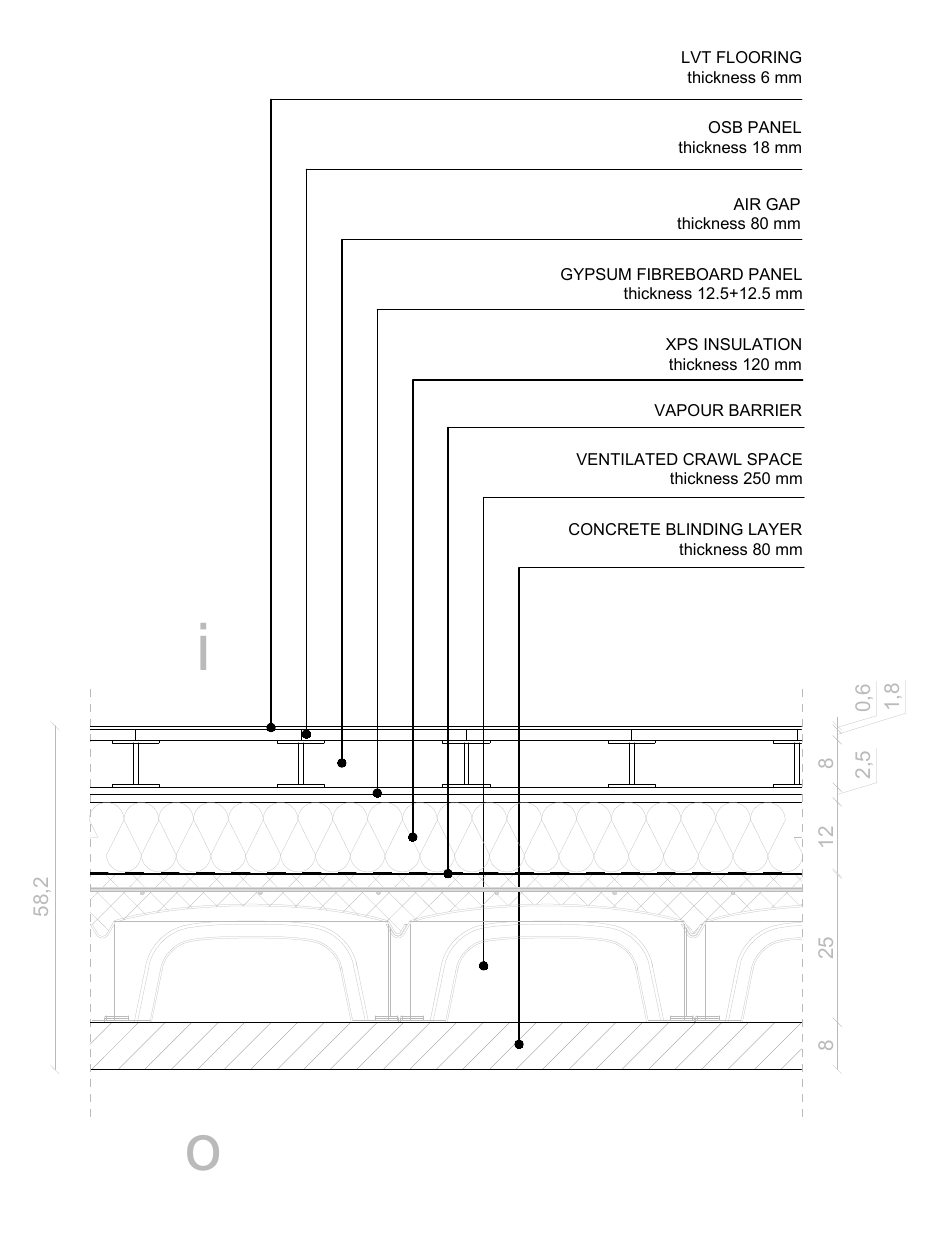}
		\caption{Horizontal enclosure against ground}
	\end{subfigure}
	\caption{Light-weight configuration stratigraphy. This Figure shows the stratigraphy of the main elements constituting the light-weight version of the building hosting the designed office room.}
	\label{fig:stratigraphy_light}
\end{figure}

\begin{figure}[!h]
	\centering
	\begin{subfigure}[b]{0.45\textwidth}
		\centering
		\includegraphics[width=\textwidth]{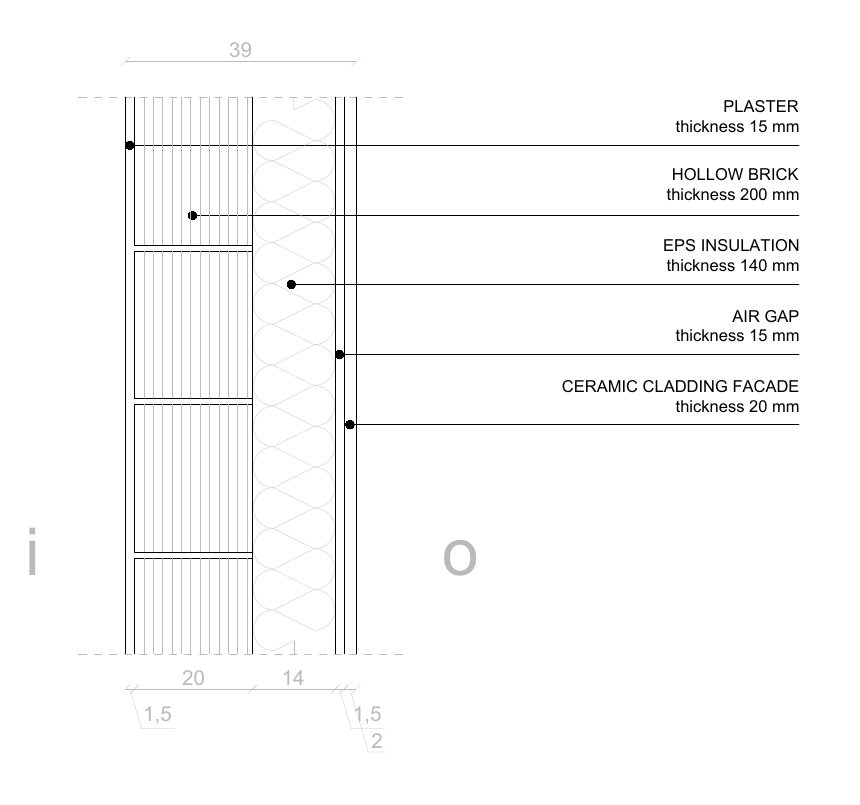}
		\caption{External opaque enclosure}
	\end{subfigure}
	\begin{subfigure}[b]{0.45\textwidth}
		\centering
		\includegraphics[width=\textwidth]{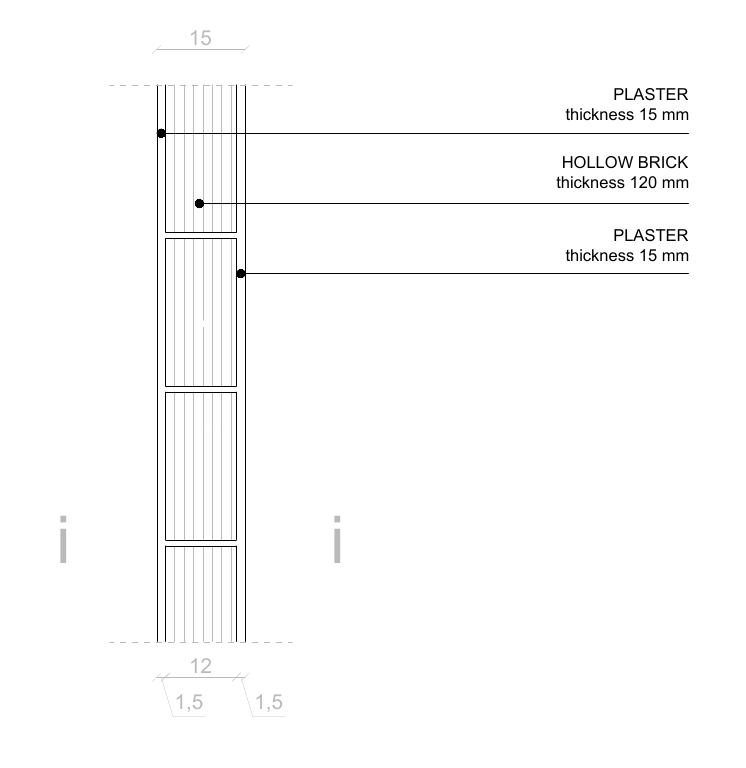}
		\caption{Internal vertical enclosure}
	\end{subfigure}
	\begin{subfigure}[b]{0.45\textwidth}
		\centering
		\includegraphics[width=\textwidth]{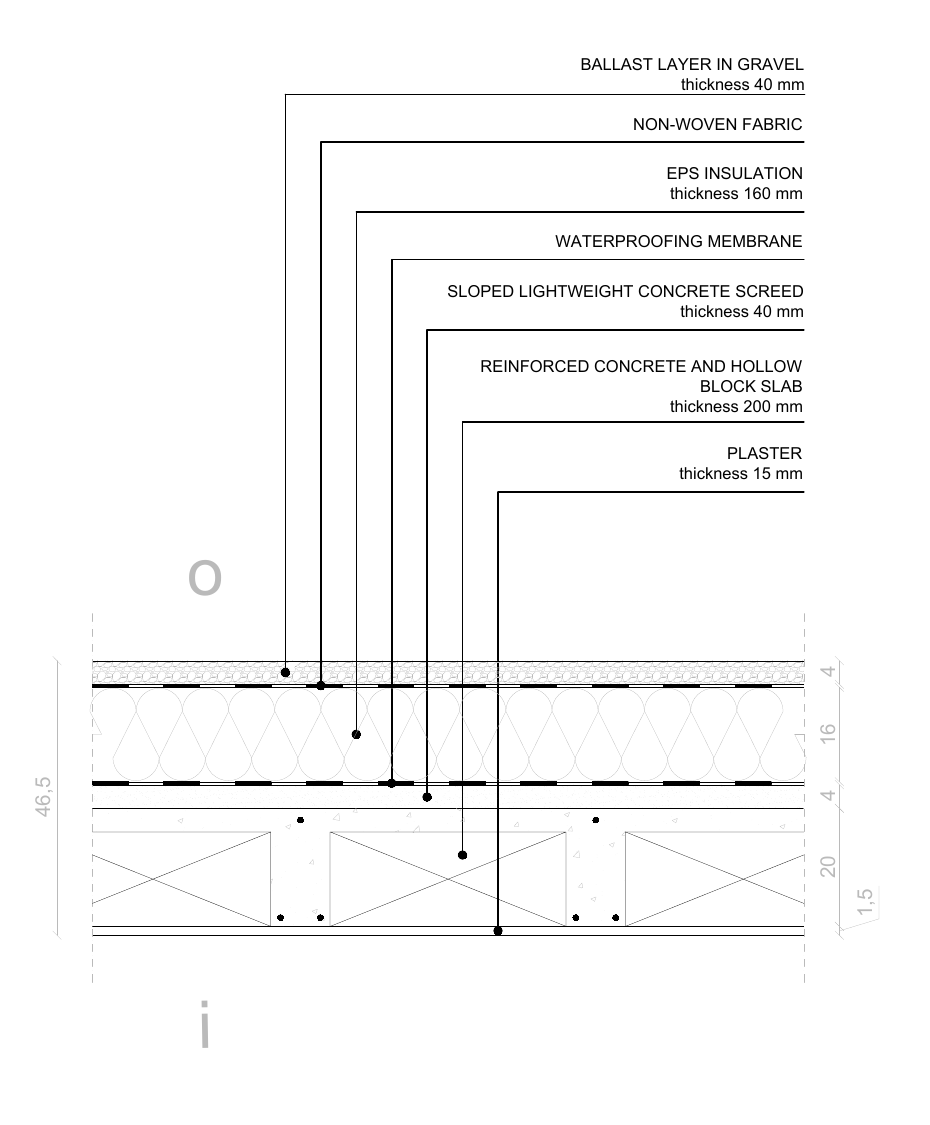}
		\caption{Horizontal flat enclosure}
	\end{subfigure}
	\begin{subfigure}[b]{0.45\textwidth}
		\centering
		\includegraphics[width=\textwidth]{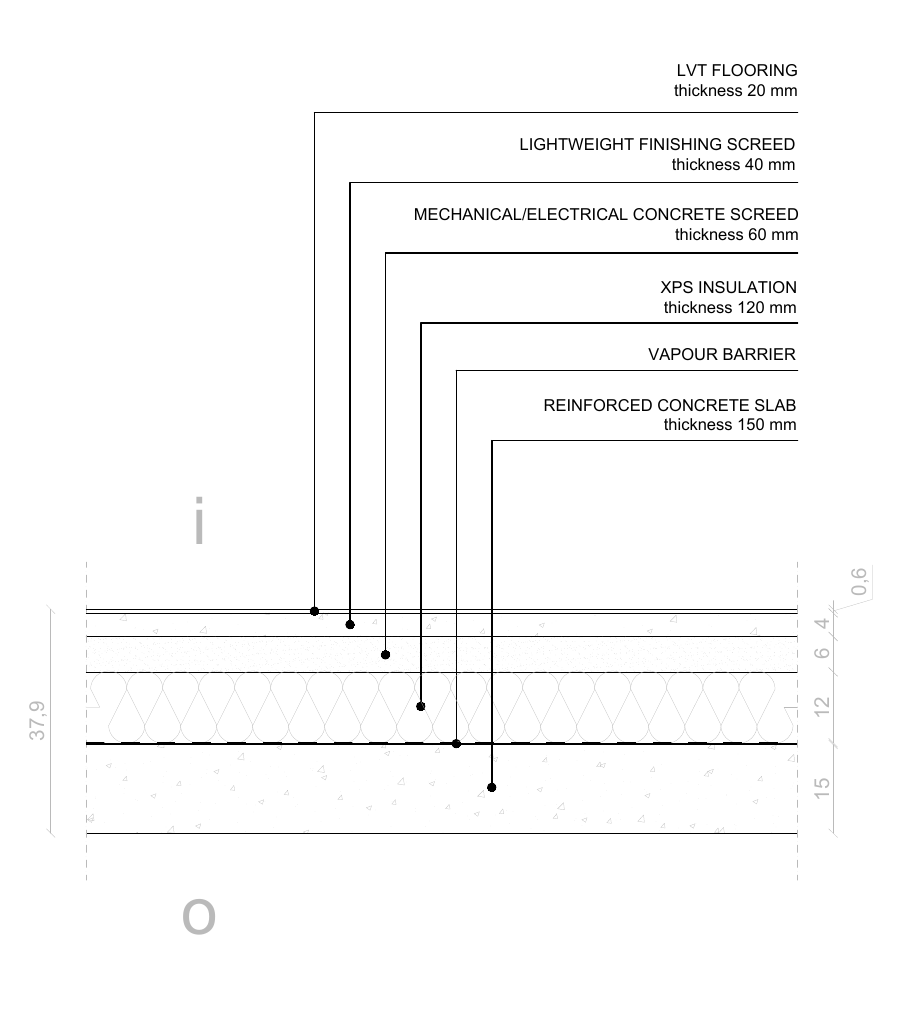}
		\caption{Horizontal enclosure against ground}
	\end{subfigure}
	\caption{Medium-weight configuration stratigraphy. This Figure shows the stratigraphy of the main elements constituting the medium-weight version of the building hosting the designed office room.}
	\label{fig:stratigraphy_medium}
\end{figure}

\begin{figure}[!h]
	\centering
	\begin{subfigure}[b]{0.45\textwidth}
		\centering
		\includegraphics[width=\textwidth]{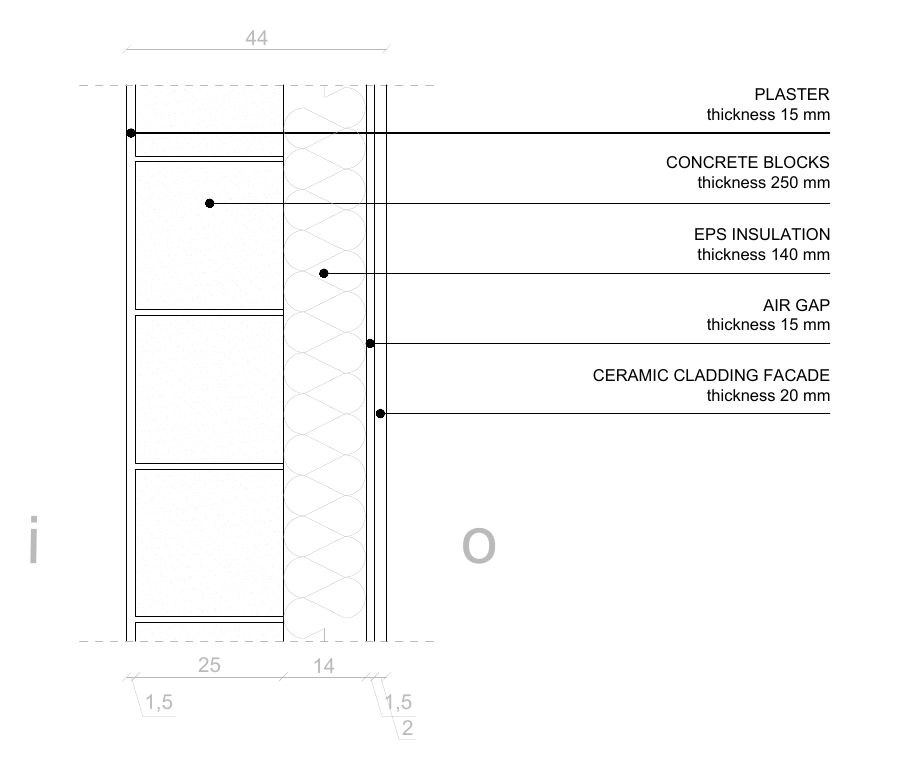}
		\caption{External opaque enclosure}
	\end{subfigure}
	\begin{subfigure}[b]{0.45\textwidth}
		\centering
		\includegraphics[width=\textwidth]{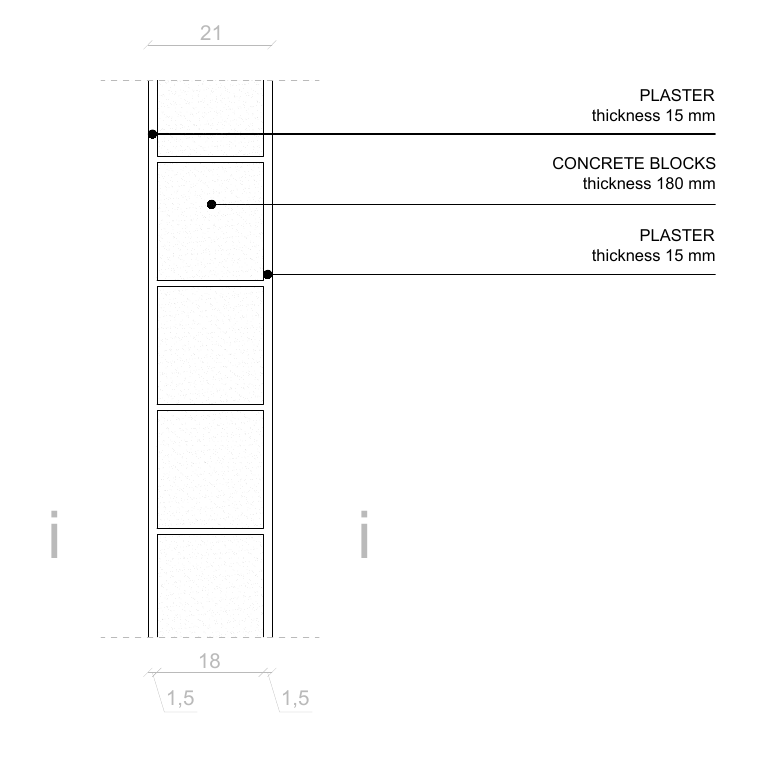}
		\caption{Internal vertical enclosure}
	\end{subfigure}
	\begin{subfigure}[b]{0.45\textwidth}
		\centering
		\includegraphics[width=\textwidth]{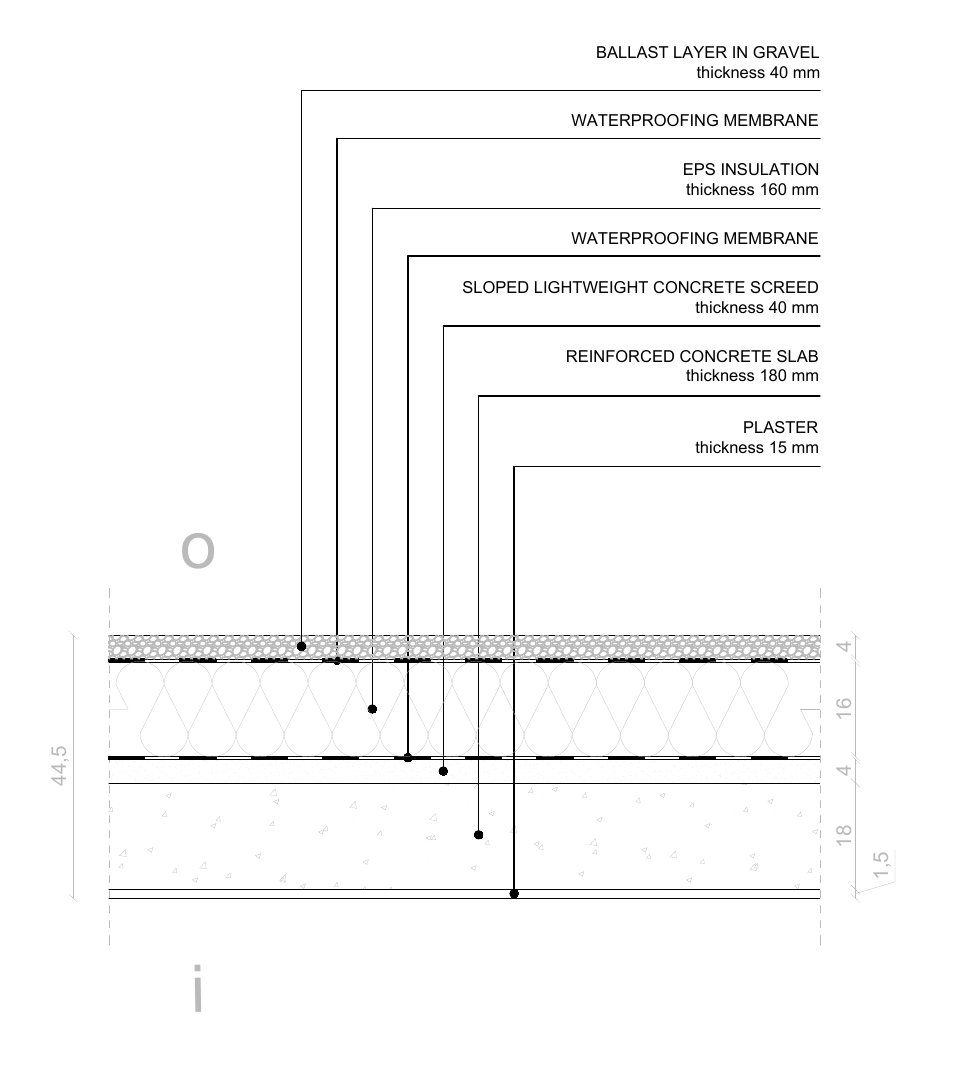}
		\caption{Horizontal flat enclosure}
	\end{subfigure}
	\begin{subfigure}[b]{0.45\textwidth}
		\centering
		\includegraphics[width=\textwidth]{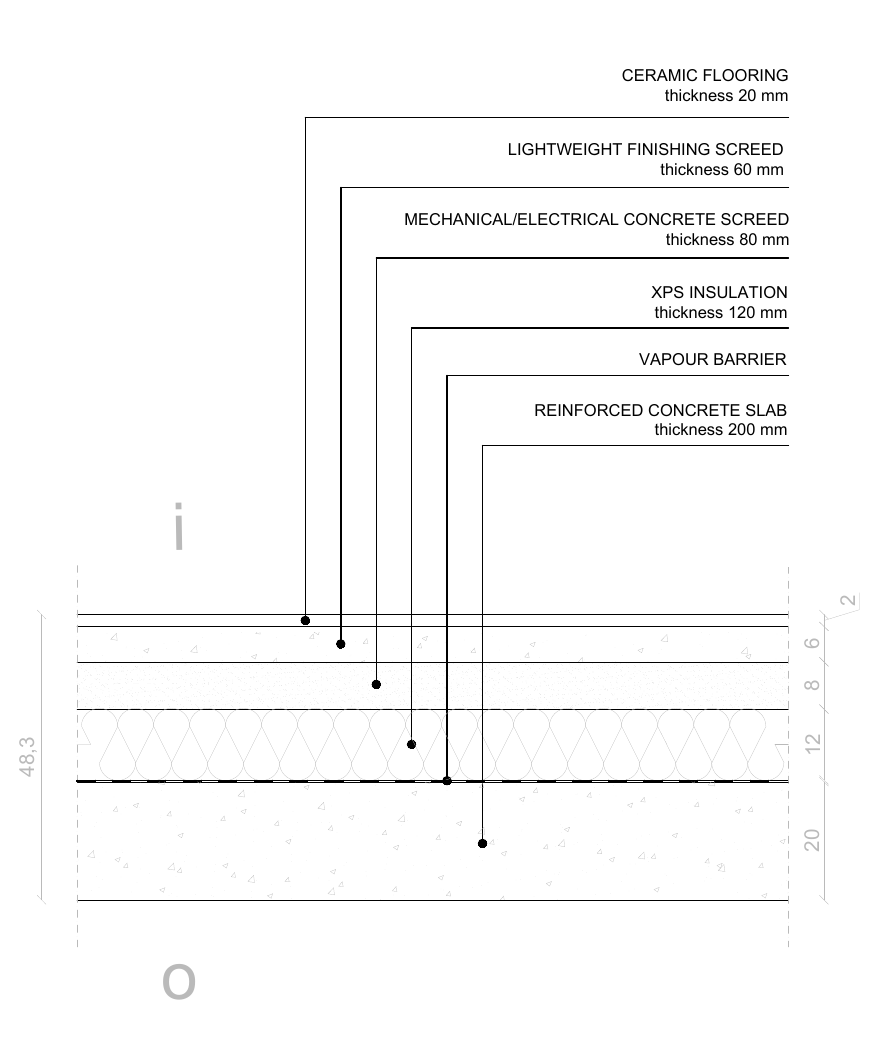}
		\caption{Horizontal enclosure against ground}
	\end{subfigure}
	\caption{Heavy-weight configuration stratigraphy. This Figure shows the stratigraphy of the main elements constituting the heavy-weight version of the building hosting the designed office room.}
	\label{fig:stratigraphy_heavy}
\end{figure}

In this Appendix, further details on the TRNSYS model implementation and on the calculation of PMV for the different scenarios are reported. First, Figure~\ref{fig:plant_building} shows the representation of the space in plan and section, carried out along the AA’ axis of the case study. In addition, Figures~\ref{fig:stratigraphy_light}, \ref{fig:stratigraphy_medium} and~\ref{fig:stratigraphy_heavy} illustrate the stratigraphy of the various envelope components used to model the office room.

Second, Table~\ref{tab:TRNSYS_info} provides further information into TRNSYS Type settings and functionality.

\begin{table}[!h]
	\caption{Employed TRNSYS Types. This Table lists all the TRNSYS Types used to design the considered office room and reports their parameter settings.}
	\raggedright
	\resizebox{.99\textwidth}{!}{
		\begin{tabular}{cc|c}
			\toprule
			\textbf{Type} & \textbf{Definition} & \textbf{Settings}\\
			\midrule
			\hline
			\multirow{4}{*}{\textbf{941}} & &  Blower Power = 0.059 $kW$\\
			& \textbf{Air To Water} & Total Air Flowrate = 1097.04 $m^3/hr$\\
			& \textbf{Heat Pump} & Rated Heating Capacity = 1.04 $kW$\\
			& & Rated Heating Power = 0.246 $kW$\\
			\hline
			\multirow{5}{*}{\textbf{954a}} & &  Total Air Flowrate = 378 $m^3/hr$\\
			& \textbf{Air to} & Rated Indoor Fan Power = 0,035 $kW$\\
			& \textbf{Heat} & Rated Outdoor Fan Power = 0,05 $kW$\\
			& \textbf{Pump} & Rated Total Cooling Capacity = 1,3\\
			& & Rated Cooling Power = 0,33 $kW$\\
			\hline
			\textbf{114} & \textbf{Circulation Pump} & Rated Power = 32 $W$\\
			\hline
			\multirow{5}{*}{\textbf{1231}} & &  Design Capacity = 1200 $W$\\
			& & Design Surface Temperature = 70 $^\circ\mathrm{C}$\\
			& \textbf{Radiator} & Design Air Temperature = 20 $^\circ\mathrm{C}$\\
			& & Design Delta-T Exponent = 1,3\\
			& & Number of Pipes = 20\\
			\hline
			\multirow{2}{*}{\textbf{562f}} & \textbf{Photovoltaic} &  Area = 7.73 $m^2$\\
			& \textbf{Panel} & PV Cell Efficiency = 23\%\\
			\hline
			\multirow{6}{*}{\textbf{23}} &  &  Minimum Control Signal = 0\\
			& & Maximum Control Signal = 1\\
			& \textbf{Cooling} & Gain Constant = -2\\
			& \textbf{Controller} & Integral Time = 1 hr\\
			& & Derivative Time = 0 hr\\
			& & Fraction of ySet for Proportional Effect = 1\\
			\hline
			\multirow{6}{*}{\textbf{23}} &  &  Minimum Control Signal = 0\\
			& \textbf{Heating} & Maximum Control Signal = 1\\
			& \textbf{and} & Gain Constant = 0.1\\
			& \textbf{Pump} & Integral Time = 1 hr\\
			& \textbf{Controller} & Derivative Time = 0 hr\\
			& & Fraction of ySet for Proportional Effect = 0.5\\
			\hline
			\bottomrule
	\end{tabular}}
	\label{tab:TRNSYS_info}
\end{table}

Last, Table~\ref{tab:PMV_info} presents additionals details on the parameters adopted for the PMV calculation across the different scenarios considered.

\begin{table}[!h]
	\caption{PMV calculation in different scenarios. The table summarizes the parameters used for the PMV calculation across the considered scenarios.}
	\raggedright
	\resizebox{.99\textwidth}{!}{
		\begin{tabular}{cc|c}
			\toprule
			\textbf{Scenario} & \textbf{Settings} & \textbf{PMV}\\
			\midrule
			\hline
			\multirow{5}{*}{\textbf{Winter scenario without $\alpha$}} & Air temperature = 20 $^\circ C$ & \\
			& Air speed = 0.1 $m/s$ & \\
			& Relative humidity  = 50\% & \textbf{PMV = -0.31}\\
			& Metabolic rate = 1.2 (70 $W/m^2$)\\
			& Clothing insulation = 1.1 (0.17 $m^2K/W$)\\
			\hline
			& Air temperature = 21.2 $^\circ C$ & \\
			& Air speed = 0.1 $m/s$ & \\
			\textbf{Winter scenario with $\alpha$} & Relative humidity  = 50\% & \textbf{PMV = -0.05}\\
			(maximum $\Delta T$ = 1.2 $^\circ C$) & Metabolic rate = 1.2 (70 $W/m^2$)\\
			& Clothing insulation = 1.1 (0.17 $m^2K/W$)\\
			\hline
			\multirow{5}{*}{\textbf{Summer scenario without $\alpha$}} & Air temperature = 26 $^\circ C$ & \\
			& Air speed = 0.15 $m/s$ & \\
			& Relative humidity  = 50\% & \textbf{PMV = 0.33}\\
			& Metabolic rate = 1.2 (70 $W/m^2$)\\
			& Clothing insulation = 0.6 (0.095 $m^2K/W$)\\
			\hline
			& Air temperature = 24.8 $^\circ C$ & \\
			& Air speed = 0.15 $m/s$ & \\
			\textbf{Summer scenario with $\alpha$} & Relative humidity  = 50\% & \textbf{PMV = -0.04}\\
			(maximum $\Delta T$ = 1.2 $^\circ C$)& Metabolic rate = 1.2 (70 $W/m^2$)\\
			& Clothing insulation = 0.6 (0.095 $m^2K/W$)\\
			\hline
			\bottomrule
	\end{tabular}}
	\label{tab:PMV_info}
\end{table}

% Bibliography
\clearpage
\bibliographystyle{elsarticle-num-names} 
\bibliography{bibliography}

\end{document}